\documentclass[fleqn,usenatbib]{mnras}

\usepackage{newtxtext,newtxmath}

\usepackage[T1]{fontenc}

\DeclareRobustCommand{\VAN}[3]{#2}
\let\VANthebibliography\thebibliography
\def\thebibliography{\DeclareRobustCommand{\VAN}[3]{##3}\VANthebibliography}

\usepackage{graphicx}	
\usepackage{amsmath}	
\usepackage{subfigure}
\usepackage{natbib}
\usepackage{color}
\usepackage{tabularx}
\usepackage{longtable}
\usepackage{bm}
\usepackage{threeparttable}
\usepackage[normalem]{ulem}

\newcommand{\code}[1]{{\texttt{#1}}}
\newcommand{\tess}{{\it TESS}}
\newcommand{\kepler}{{\it Kepler}}

\newcommand{\gaia}{{\it Gaia}}
\newcommand{\jwst}{{\it JWST}}
\newcommand{\elt}{{\it ELT}}
\newcommand{\tar}{{HD 183579}}
\newcommand{\flatiron}{Center for Computational Astrophysics, Flatiron Institute, 162 Fifth Ave, New York, NY 10010, USA}
\newcommand{\chicago}{Department of Astronomy and Astrophysics, University of Chicago, 5640 S. Ellis Ave, Chicago, IL 60637, USA}
\newcommand{\USP}{Departamento de Astronomia, IAG, Universidade de S{\~a}o Paulo, Rua do Mat{\~a}o 1226, S{\~a}o Paulo, 05509-900, Brazil}
\newcommand{\MIT}{Department of Physics and Kavli Institute for Astrophysics and Space Research, Massachusetts Institute of Technology, Cambridge, MA 02139, USA}
\newcommand{\harvard}{Center for Astrophysics ${\rm \mid}$ Harvard {\rm \&} Smithsonian, 60 Garden Street, Cambridge, MA 02138, USA}

\title[TOI-1055b]{HD 183579b: A Warm Sub-Neptune Transiting a Solar Twin Detected by TESS}

\author[T. Gan et al.]{Tianjun Gan,$^{1}$\thanks{E-mail: gtj18@mails.tsinghua.edu.cn}
Megan Bedell,$^{2}$
Sharon~Xuesong~Wang,$^{1}$
Daniel Foreman-Mackey,$^{2}$
Jorge Mel{\'e}ndez,$^{3}$
\newauthor
Shude~Mao,$^{1,4}$
Keivan G.\ Stassun,$^{5,6}$
Steve~B.~Howell,$^{7}$
Carl Ziegler,$^{8}$
Robert A. Wittenmyer,$^{9}$
\newauthor
Coel~Hellier,$^{10}$
Karen A.\ Collins,$^{11}$
Avi Shporer,$^{12}$
George~R.~Ricker,$^{12}$
Roland~Vanderspek,$^{12}$
\newauthor
David~W.~Latham,$^{11}$
Sara~Seager,$^{12,13,14}$
Joshua~N.~Winn,$^{15}$
Jon~M.~Jenkins,$^{7}$
Brett~C.~Addison,$^{9}$
\newauthor
Sarah Ballard,$^{16}$
Thomas~Barclay,$^{17,18}$
Jacob L. Bean,$^{19}$
Brendan P. Bowler,$^{20}$
C\'{e}sar Brice\~{n}o,$^{21}$
\newauthor
Ian~J.~M.~Crossfield,$^{22}$
Jason Dittman,$^{11,23}$
Jonathan Horner,$^{9}$
Eric L.\ N.\ Jensen,$^{24}$
Stephen R. Kane,$^{25}$
\newauthor
John Kielkopf,$^{26}$
Laura Kreidberg,$^{11,23}$
Nicholas Law,$^{27}$
Andrew W. Mann,$^{27}$
Matthew W. Mengel,$^{9}$
\newauthor
Edward~H.~Morgan,$^{12}$
Jack Okumura,$^{9}$
Hugh~P.~Osborn,$^{12,28}$
Martin~Paegert,$^{11}$
Peter Plavchan,$^{29}$
\newauthor
Richard P. Schwarz, $^{30}$
Bernie Shiao,$^{31}$
Jeffrey~C.~Smith,$^{7,32}$
Lorenzo Spina,$^{33}$
C.~G.~Tinney,$^{34}$
\newauthor
Guillermo~Torres,$^{11}$
Joseph D. Twicken,$^{7,32}$
Michael~Vezie,$^{12}$
Gavin Wang,$^{35,36}$
Duncan J. Wright,$^{9}$
\newauthor
and Hui Zhang$^{37}$
\\
Affiliations are listed at the end of the paper
}

\date{Accepted XXX. Received YYY; in original form ZZZ}

\pubyear{2021}

\begin{document}
\label{firstpage}
\pagerange{\pageref{firstpage}--\pageref{lastpage}}
\maketitle

\begin{abstract}
We report the discovery and characterization of a transiting warm sub-Neptune planet around the nearby bright ($V=8.75$~mag, $K=7.15$~mag) solar twin \tar, delivered by the Transiting Exoplanet Survey Satellite (\tess). The host star is located $56.8\pm0.1$ pc away with a radius of $R_{\ast}=0.97\pm0.02\ R_{\odot}$ and a mass of $M_{\ast}=1.03\pm0.05\ M_{\odot}$. We confirm the planetary nature by combining space and ground-based photometry, spectroscopy, and imaging. We find that \tar b (TOI-1055b) has a radius of $R_{p}=3.53\pm0.13\ R_{\oplus}$ on a $17.47$~day orbit with a mass of $M_{p}=11.2\pm5.4\ M_{\oplus}$ ($3\sigma$ mass upper limit of $27.4\ M_{\oplus}$). \tar b is the fifth brightest known sub-Neptune planet system in the sky, making it an excellent target for future studies of the interior structure and atmospheric properties. By performing a line-by-line differential analysis using the high resolution and signal-to-noise ratio HARPS spectra, we find that \tar\ joins the typical solar twin sample, without a statistically significant refractory element depletion. 
\end{abstract}




\section{Introduction}
After the first discovery of a hot Jupiter outside our Solar system \citep{Mayor1995}, exoplanet research has moved into a new era. Up to now, more than 4\,000 exoplanets have been confirmed \footnote{\url{https://exoplanetarchive.ipac.caltech.edu/}}. Most giant planets have been found by successful ground surveys like HATNet \citep{Bakos2004}, SuperWASP \citep{Pollacco2006}, KELT \citep{Pepper2007,Pepper2012} and NGTS \citep{Chazelas2012,Wheatley2018}. Space mission conducting photometric transit surveys including {\it CoRoT} \citep{Baglin2006}, {\it Kepler} \citep{Borucki2010} and {\it K2} \citep{Howell2014} have led to the further detections of thousands of planets with size between Earth and Neptune. These diverse exoplanets are hosted by a similarly diverse set of stars. Among them, Sun-like stars (here defined as FGK main-sequence stars) make up a significant fraction of known planet hosts. These systems can be seen as an intriguing opportunity to get a glimpse into alternate paths to that our own Solar system might have taken in its early formation, and they represent our best opportunity to discover a ``truly Earth-like'' exoplanet that exists under conditions as similar as possible to our own planet \citep{Horner2020,Kane2020}.


Solar twins are an important subset of Sun-like stars. Typically defined by their extreme similarity to the Sun in fundamental spectroscopic properties ($T_{\rm eff}$ within 100 K, $\log g$ within 0.1 dex, and [Fe/H] within 0.1 dex of Solar values), these stars must by definition have such similar photospheric conditions to the Sun so that their spectra can be directly compared with minimal reliance on stellar atmospheric models. The result of a line-by-line differential spectroscopic analysis of a solar twin yields uniquely precise abundance measurements for the star and thereby for the star-planet system \citep[see e.g.][who achieve 0.01 dex or 2\% precision on abundance measurements for over 30 elements]{Bedell2018,Spina2018}. This is in direct contrast to a typical planet host star, whose abundances are expected to be limited by systematic uncertainties to the level of 0.05 dex or more. Similarly precise measurements may be made of the star's age, mass, radius, and other fundamental properties by combining isochronal models with the spectroscopic measurements \citep{Ramirez2014,Yana2016}. It is worth emphasizing that these properties are measured with extreme precision (not necessarily accuracy) relative to the Sun, our most thoroughly characterized planet host. Planetary systems around solar twin stars are therefore useful both as individual well-characterized planets but also as a prime sample for comparative studies delving into any subtle differences between stars that host planets of different types. Unfortunately, the sample of solar twins with well characterized planets around is still limited in number at present (e.g., \kepler-11, \citealt{Lissauer2011}; HIP 11915, \citealt{Bedell2015}; K2-231, \citealt{Curtis2018}; KELT-22, \citealt{Labadie2019}), roughly 50 in total. 

The Transiting Exoplanet Survey Satellite (\tess, \citealt{Ricker2014,Ricker2015}), which performs an all-sky survey and focuses on small exoplanets orbiting nearby bright stars, will likely increase the sample of planets around solar twins significantly \citep{Sullivan2015,Huang2018}. During its two-year primary mission, \tess\ has detected over two thousand exoplanet candidates, the majority of which are suitable for follow-up observations, including mass measurements and atmospheric spectroscopy. This makes \tess\ planet candidates unlike most \kepler\ systems, which are too faint for these follow-up observation.

In this work, we present a warm sub-Neptune planet detected by \tess\ to orbit a solar twin star \tar. \tar\ is a G2V star with a spectrum nearly identical to that of the Sun. The star has been studied extensively through a dedicated RV planet search and spectroscopic abundance survey targeting solar twin stars with the High Accuracy Radial velocity Planet Searcher spectrograph (HARPS, \citealt{Mayor2003}; \citealt{Melendez2015}). The transiting planet, however, was not detected until \tess\ data became available. 

This paper is organized as follows: In Section \ref{observations}, we describe all observations. We characterize the host star \tar\ in Section \ref{stellar_properties}. Section \ref{analysis} presents our analysis of the light curves and RV data. The lessons about comparison between \tar\ and other similar systems are discussed in Section \ref{solar_analogs}. In Section \ref{discussion}, we discuss insights into this system, prospects for further characterization via transmission spectroscopy, a search for additional planets, and a comparison with a recently published analysis of archival RVs of this target \citep{Palatnick2021}. We conclude our findings in Section \ref{conclusion}.


\section{Observations}\label{observations}
\subsection{\tess}
\tar\ (TIC 320004517) was monitored by \tess\ with the two-minute cadence mode in Sector 13 during the primary mission and Sector 27 during the extended mission. The data were obtained between 2019 June 19th and 2019 July 18th, and between 2020 July 5th and 2020 July 30th, consisting of a total of 20479 and 17546 individual measurements, respectively. 

The raw images were reduced using the Science Processing Operations Center\ (SPOC) pipeline \citep{Jenkins2016}, which was developed at NASA Ames Research Center based on the \kepler\ mission's science pipeline. After the systematic and dilution effects were corrected by the Presearch Data Conditioning (PDC; \citealt{Stumpe2012,Smith2012,Stumpe2014}) module, Transiting Planet Search \citep[TPS;][]{Jenkins2002,Jenkins2017} was then performed to look for transit-like signals. \tar\ was finally identified as a planet candidate in the \tess\ Object of Interest catalog (TOI 1055.01) with a period of 17.47 days and a transit depth of 1259 ppm \citep{Twicken2018,Li2019}, and alerted on the MIT \tess\ Alerts portal\footnote{\url{https://tess.mit.edu/alerts/}}.

We downloaded the Presearch Data Conditioning Simple Aperture Photometry (PDCSAP) light curve from the Mikulski Archive for Space Telescopes\ (MAST\footnote{\url{http://archive.stsci.edu/tess/}}). After removing all measurements flagged for quality issues in SPOC to improve the precision, we applied the built-in routines of the \code{lightkurve} package \citep{lightkurvecollaboration,lightkurve} to normalize the data and clip outliers above a $+5\sigma$ limit. These additional processing steps removed 904 and 802 points (4.4\% and 4.6\%), with 19575 and 16744 measurements left for each sector.

To search for potential additional planets, we smoothed the light curve with a median filter and performed an independent transit search using the Box Least Square (BLS; \citealt{Kovacs2002}) algorithm. We confirmed the $\sim 17$ d signal reported by TPS. Except for that, we did not detect any other significant peaks existing in the periodogram. 

\begin{figure*}
\centering
\includegraphics[width=\textwidth]{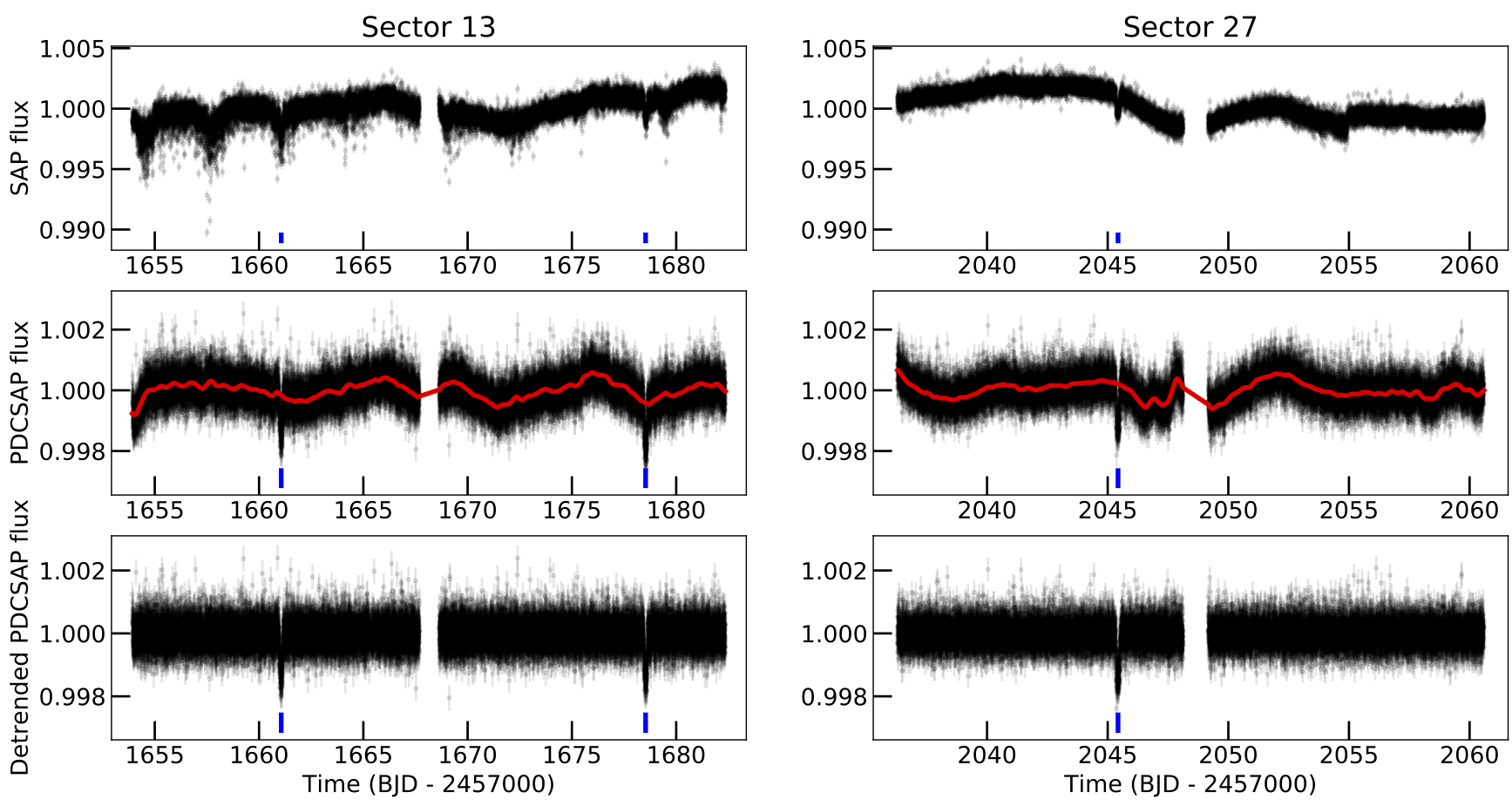}
\caption{{\it Top panels}: The original \tess\ SAP light curves of \tar\ from Sector 13 and 27. {\it Middle panels}: The PDCSAP light curves of \tar\ along with the best-fit GP model shown as red solid lines. {\it Bottom panels}: The detrended PDCSAP light curves. The three transits of \tar b are marked in blue ticks.} 
\label{transit_detrend}
\end{figure*}

After masking out all in-transit data, we detrended the light curve by fitting a Gaussian Process (GP) model with a simple Matern32 kernel using the \code{celerite} package \citep{Foreman2017}. Figure \ref{transit_detrend} shows the original SAP, PDCSAP and the PDCSAP light curve after detrending. We used this reprocessed light curve in our further transit analysis.

\subsection{Ground-Based Photometry}\label{gbp}
\subsubsection{Las Cumbres Observatory (LCO)}\label{lco}
\begin{figure*}
\centering
\includegraphics[width=0.95\columnwidth]{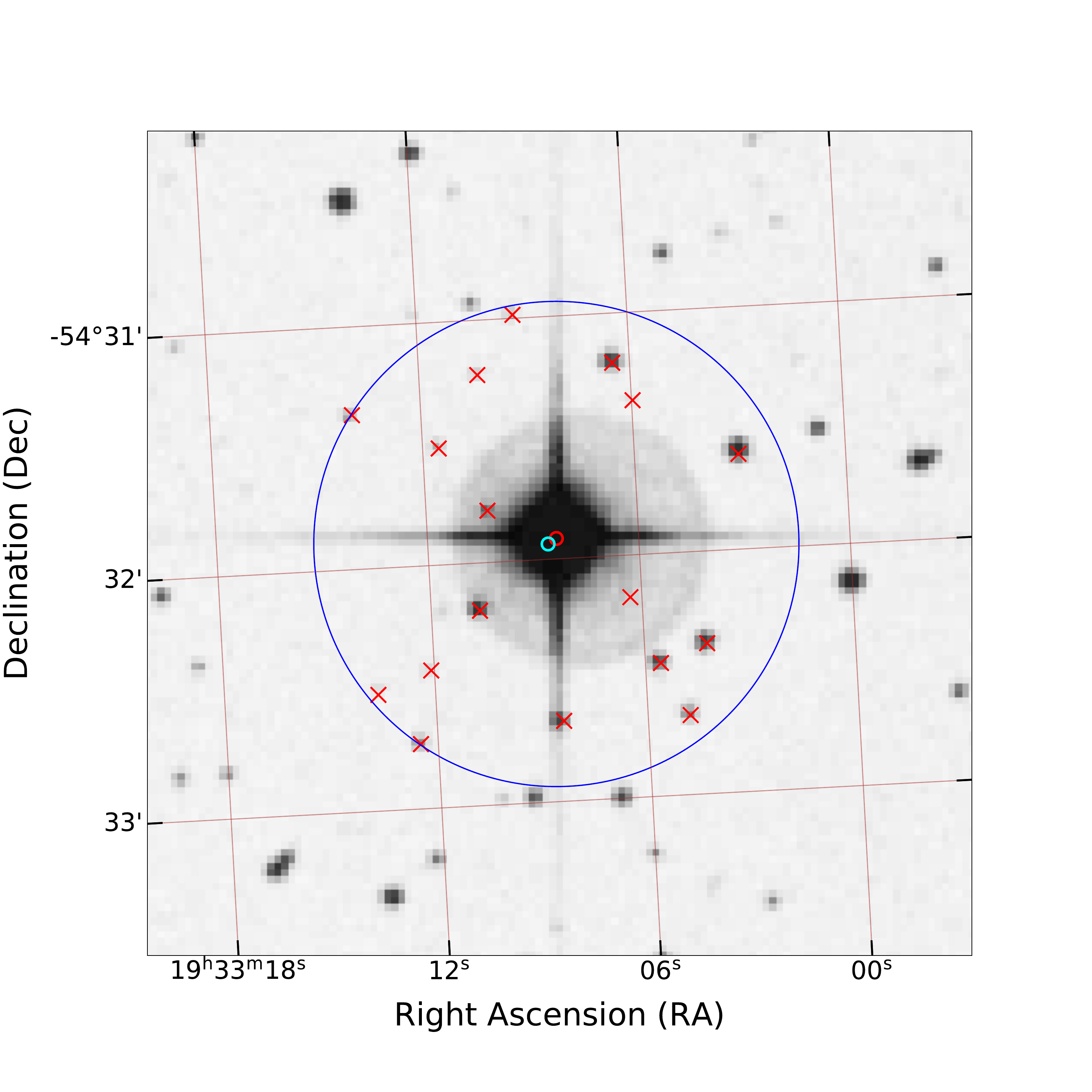}
\includegraphics[width=1.05\columnwidth]{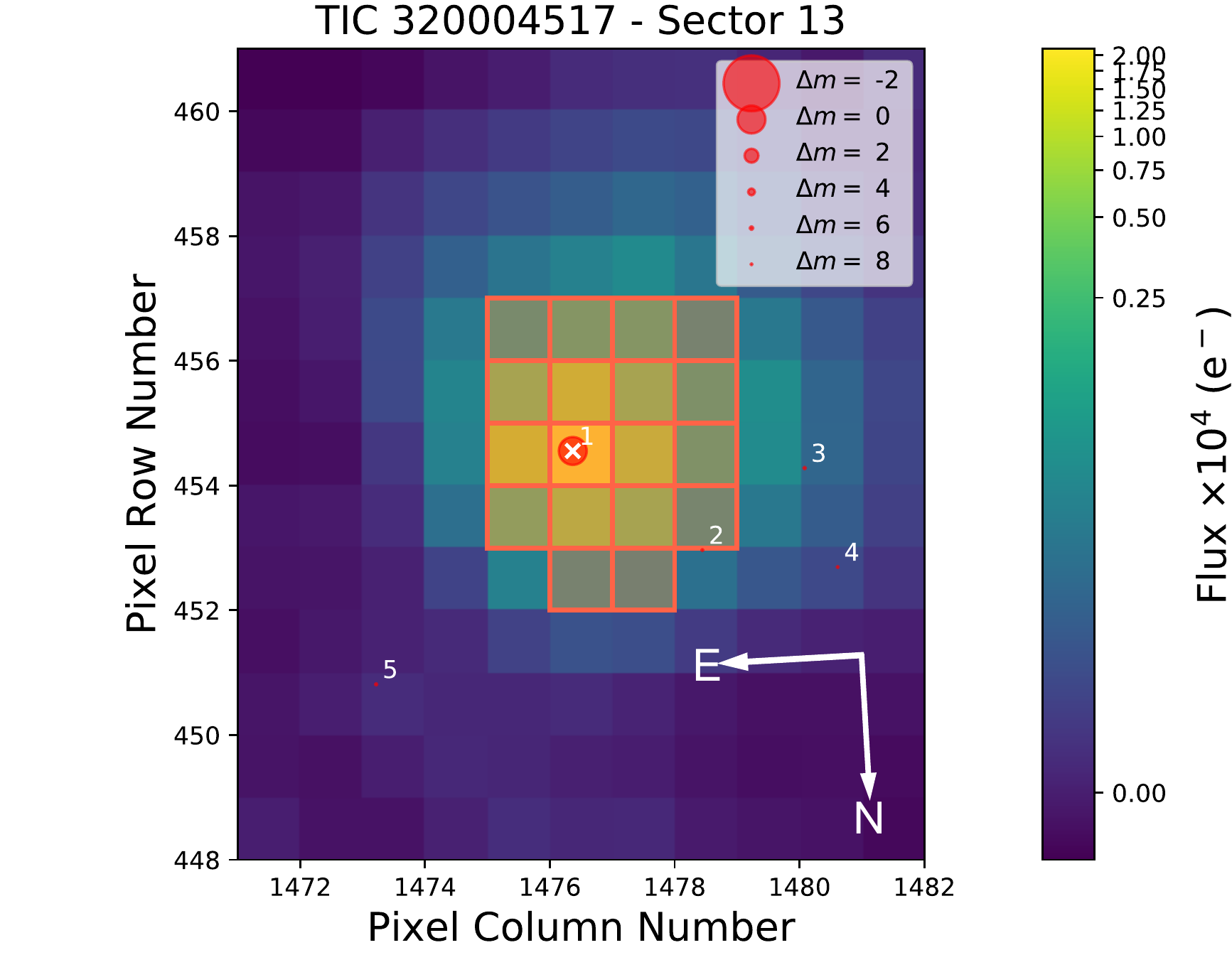}
\caption{{\it Left panel}: The POSS2 blue image of \tar\ taken in 1976. The center red dot is the target star in this image and the cyan dot shows its current position. All stars (marked as red crosses) in 1$'$ (the blue circle) are ruled out as the source that causes the \tess\ detection based on their brightness and the NEB analysis of LCO photometry. {\it Right panel}: Target pixel file (TPF) of \tar\ in \tess\ Sector 13 (created with \code{tpfplotter}, \citet{Aller2020}). Different sizes of red circles represent different magnitudes in contrast with \tar\ ($\Delta m$). The red-square region represents the aperture used to extract the photometry by SPOC. The light contamination from nearby stars is negligible (see Section \ref{transit}).}
\label{fov}
\end{figure*}

The large pixel scale of \tess\ ($21''$ per pixel, \citealt{Ricker2014,Ricker2015}) may result in light contamination from stars close to the target, making nearby eclipsing binaries\ (NEB) a common source of \tess\ false positives \citep{Brown2003,Sullivan2015}. To rule out the NEB scenario and confirm the event on target, we collected two ground-based follow-up observations using the Las Cumbres Observatory Global Telescope\ (LCOGT\footnote{\url{https://lco.global/}}) network \citep{Brown2013}. We used the \tess\ Transit Finder (\code{TTF}), which is a customized version of the \code{Tapir} software package \citep{Jensen2013}, to schedule these time-series observations. The photometric observations were taken in the Pan-STARRS $Y$ band with an exposure time of 35 s on 2020 June 27th and 2020 August 1st at Siding Spring Observatory\ (SSO), Australia and both were done with 1m telescopes. The Sinistro cameras have a $26' \times 26'$ field of view as well as a plate scale of $\rm 0.389''$ per pixel. The images were defocused and have stellar point-spread-functions (PSF) with a full-width-half-maximum (FWHM) of $\sim 2\farcs 4$ and $\sim 2\farcs 0$, respectively. After the images were calibrated by the standard automatic \code{BANZAI} pipeline \citep{McCully2018}, we carried out photometric analysis using \code{AstroImageJ} (\citealt{Collins2017}). We excluded all nearby stars within 1 arcmin as the source causing the \tess\ signal with brightness difference down to $ \Delta T \sim 7.5 $ mag (see Figure \ref{fov}), and confirmed the signal on target. We summarize the observations in Table \ref{po}.

\begin{table*}
    \centering
    \caption{Summary of ground-based photometric observations for \tar}
    \begin{tabular}{clccccc}
        \hline\hline
        Facility       &Date &Total exposures &Exposure time(s) &Filter  &Coverage  &Label \\\hline
        LCO 1m SSO Sinistro &2020 June 27 &224 &35 &Pan-STARRS $Y$ &egress &LCOA\\
        LCO 1m SSO Sinistro &2020 Aug 1 &274 &35 &Pan-STARRS $Y$ &egress &LCOB\\
         \hline
    \end{tabular}
    \label{po}
\end{table*}

\subsubsection{WASP}
WASP-South, an array of 8 cameras, was the Southern half of the WASP transit-search survey \citep{Pollacco2006}. The field of \tar\ was observed in both 2013 and 2014, covering a span of 180 nights in each year with a typical 10-min cadence on clear nights, and accumulating 52\,000 data points. WASP-South was equipped with 85-mm, f/1.2 lenses giving a photometric extraction aperture with a 112-arcsec radius. All other stars within this aperture are $>$\,5 mag fainter. 


\subsection{High Resolution Spectroscopy}
\subsubsection{HARPS}\label{harps}
\tar\ was observed 56 times by the High Accuracy Radial velocity Planet Searcher\ (HARPS; \citealt{Mayor2003}) on the ESO\,3.6\,m telescope at La Silla Observatory in Chile between 2011 and 2019. The bulk of these observations were made as part of a dedicated blind planet search targeting solar twins (P.I. Mel\'endez). 
All observations were carried out in high-accuracy mode with a spectral resolution R~$\sim$~115,000. The median SNR is 108 pix$^{-1}$ at 600 nm.


We extracted the radial velocity (RV) measurements, chromatic RV index (CRX) and differential line width (dLW) using the publicly available SpEctrum Radial Velocity AnaLyser pipeline (\code{SERVAL}, \citealt{Zechmeister2018}). Additional diagnostics including the inverse bisector span (BIS) and full width at half-maximum (FWHM) for the line profile of the average spectral features were extracted by the standard HARPS pipeline using a cross-correlation technique with a solar-type mask \citep{Pepe2002}. These diagnostics are commonly used as stellar activity tracers, since they quantify the line distortions which mimic Doppler shifts. 

In addition to these activity indicators, we also derived the $S_{\rm HK}$ measurement, which quantifies the strength of emission in the cores of the Ca~II~H\&K lines. These were measured and corrected to the standard Mount Wilson scaling using the procedure outlined in \cite{Lovis2011}. Measured $S_{\rm HK}$ values and photon-noise-based uncertainties along with the pipeline values of RV, BIS, FWHM, CRX and dLW are publicly available on ExoFOP-TESS\footnote{\url{https://exofop.ipac.caltech.edu/tess/target.php?id=320004517}}.

We dropped one observation (BJD=2457588.767) from the analysis because its BIS and FWHM measurements were significant outliers ($>5\sigma$) from the general distribution, pointing to potential issues with the data reduction and RV extraction.

\subsubsection{\textsc{\textsc{Minerva}}-Australis}
\textsc{\textsc{Minerva}}-Australis is an array of four PlaneWave CDK700 telescopes located at the Mt Kent Observatory in Queensland, Australia, fully dedicated to the precise radial-velocity follow-up of \tess\ candidates \citep[e.g.][]{toi677,toi257,AUMicBrett}. The four telescopes can be simultaneously fiber-fed to a single KiwiSpec R4-100 high-resolution (R=80,000) spectrograph \citep{Barnes12,addison19}. \tar\ was monitored by \textsc{\textsc{Minerva}}-Australis using up to 4 telescopes in the array between 2020 April 19 and 2020 June 1.  Each epoch consists of one or two 30-minute exposures.  Telescopes 1, 3, 4, and 5 (denoted as MA, MB, MC and MD) obtained 5, 8, 15, and 5 epochs respectively. Radial velocities for the observations are derived for each telescope by cross-correlation, where the template being matched is the mean spectrum of each telescope. A simultaneous quartz-illuminated iodine cell in the calibration fibres 
provides the wavelength calibration and corrects for instrumental 
variations. We converted all time-stamps of our measurements from JD to BJD using \code{barycorr} \citep{Wright2014}.

\subsection{High Angular Resolution Imaging}
High-angular resolution imaging is needed to search for nearby sources that can contaminate the \tess\ photometry, resulting in an underestimated planetary radius, or other sources of astrophysical false positives, such as background eclipsing binaries. 

\subsubsection{Gemini-South}
We observed \tar\ to probe for companion stars on 12 September 2019 UT using the Zorro instrument mounted on the 8\ m Gemini South telescope, located on Cerro Pach\'on in Chile. Zorro uses speckle imaging to simultaneously observe diffraction-limited images at 562 nm (0.017$''$) and 832 nm (0.028$''$). Our data set consisted of three $1000\times60$ ms exposure images simultaneously obtained in both band-passes, followed by a single $1000\times60$ ms image, also in both band-passes, of a PSF standard star.

Following the procedures outlined in \cite{Howell2011}, we combined all images and subjected them to Fourier analysis, and produce re-constructed imagery from which 5-sigma contrast curves are derived in each passband (Figure \ref{zspeckle}). Our data reveal \tar\ to be a single star to contrast limits of 5 to 8 magnitudes within the spatial limits of 1.0/1.6 AU (562/832 nm respectively) out to 57 AU.

\subsubsection{SOAR}
\begin{figure*}
\centering
\includegraphics[width=\columnwidth]{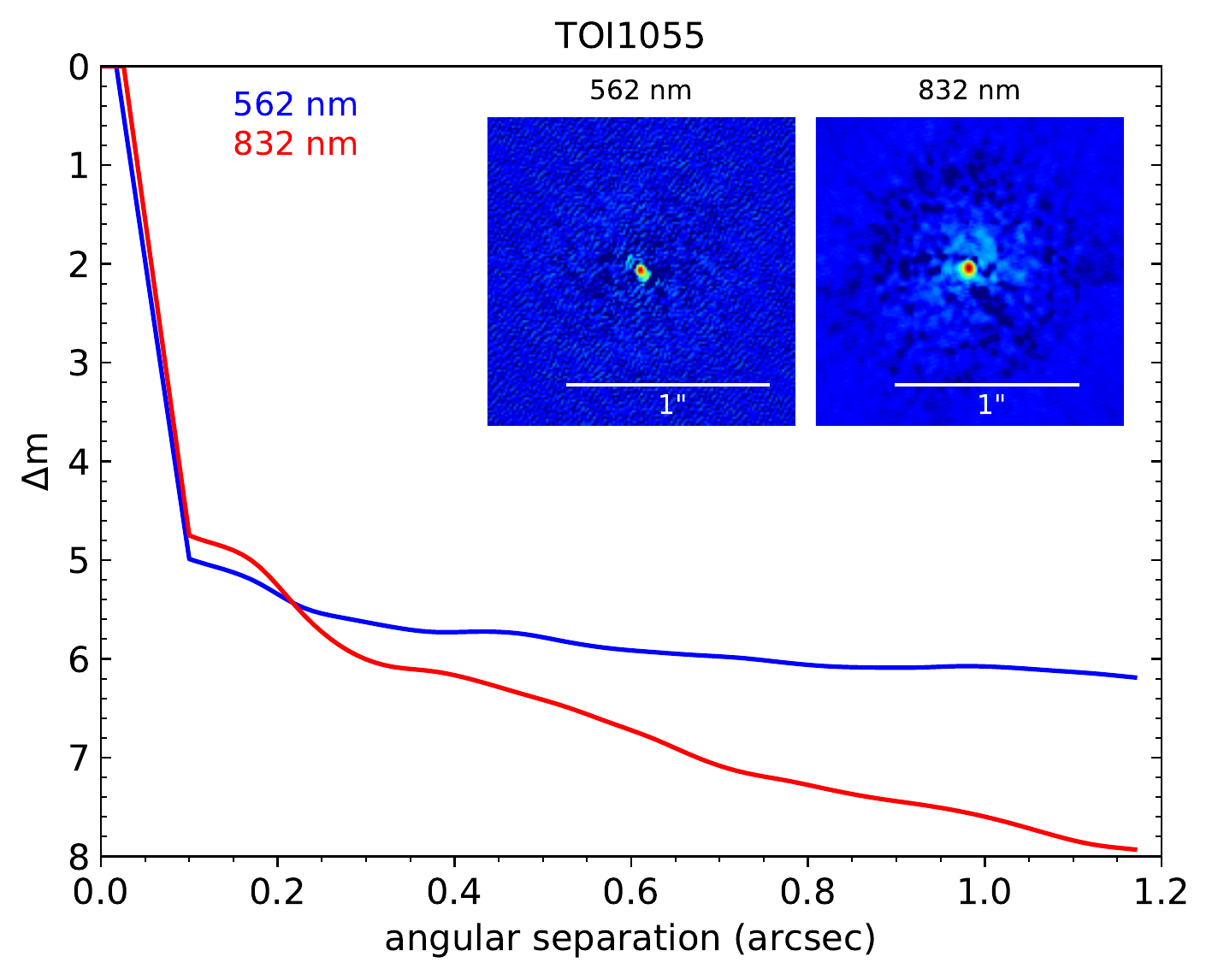}
\includegraphics[width=\columnwidth]{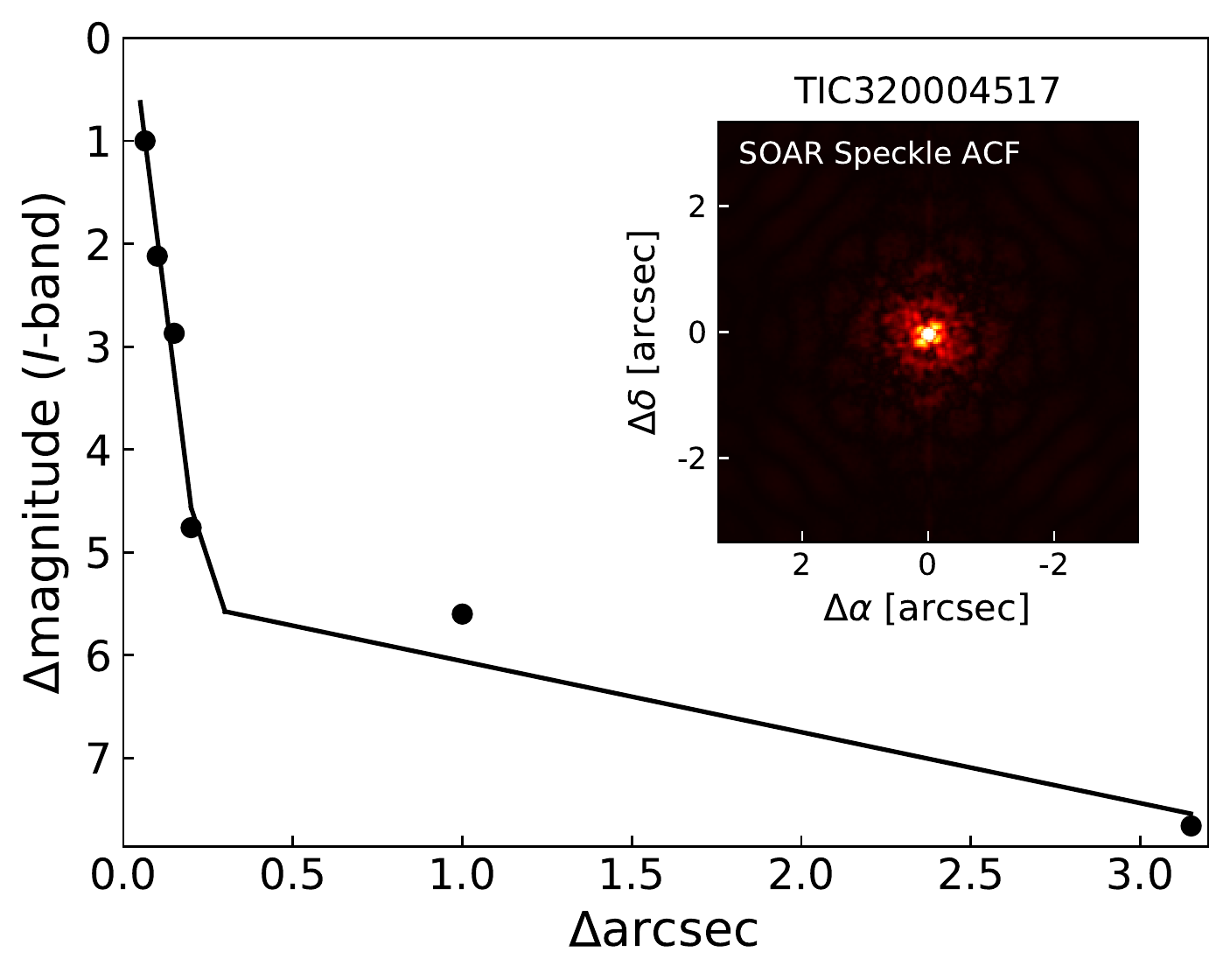}
\caption{{\it Left panel}: Zorro speckle imaging and $5\sigma$ contrast curves of \tar\ at 562\,nm and 832\,nm. The data reveal that no companion star is detected within the spatial limits of 1 AU out to 57 AU with a $\Delta$\,m of 5 to 8. {\it Right panel}: Speckle ACF obtained in the I band using SOAR. The 5$\sigma$ contrast curve for \tar\ is shown by the black points. Black solid line corresponds to the linear fit of the data, at separations smaller and larger than $\sim 0.2''$.}
\label{zspeckle}
\end{figure*}

We also searched for stellar companions to \tar\ with speckle imaging on the 4.1-m Southern Astrophysical Research (SOAR) telescope \citep{Tokovinin2018} on 31 October 2020 UT, observing in Cousins I-band, a similar visible bandpass as \tess. More details of the observations are available in \cite{Ziegler2020}. The 5$\sigma$ detection sensitivity and speckle auto-correlation functions from the observations are shown in the right panel of Figure \ref{zspeckle}. No nearby stars were detected within 3\arcsec of \tar\ in the SOAR observations.

\section{Stellar properties}\label{stellar_properties}
\subsection{Stellar Characterization}\label{stellar_characterization}

We first derived $T_{\rm eff}$, $R_{\ast}$, and iron abundance [Fe/H] from the spectroscopic data. By utilizing the \code{SpecMatch-Emp} package \citep{Yee2017}, we matched the co-added HARPS spectrum to a high resolution spectroscopic library, which contains 404 well-characterized stars, following \cite{Hirano2018}. We found $T_{\rm eff}=5678\pm110$ K, $R_{\ast}=0.988\pm0.100\ R_{\odot}$ and ${\rm [Fe/H]}=-0.07\pm0.09$ dex.
This is in good agreement with the literature values of $T_{\rm eff}=5798\pm4$ K, $\log g=4.480\pm0.012$ dex, and ${\rm [Fe/H]}=-0.036\pm0.003$ dex, as derived by \citet{Spina2018} using the same co-added HARPS observations with a strictly differential line-by-line equivalent width technique.

\begin{figure}
\centering
\includegraphics[width=0.5\textwidth]{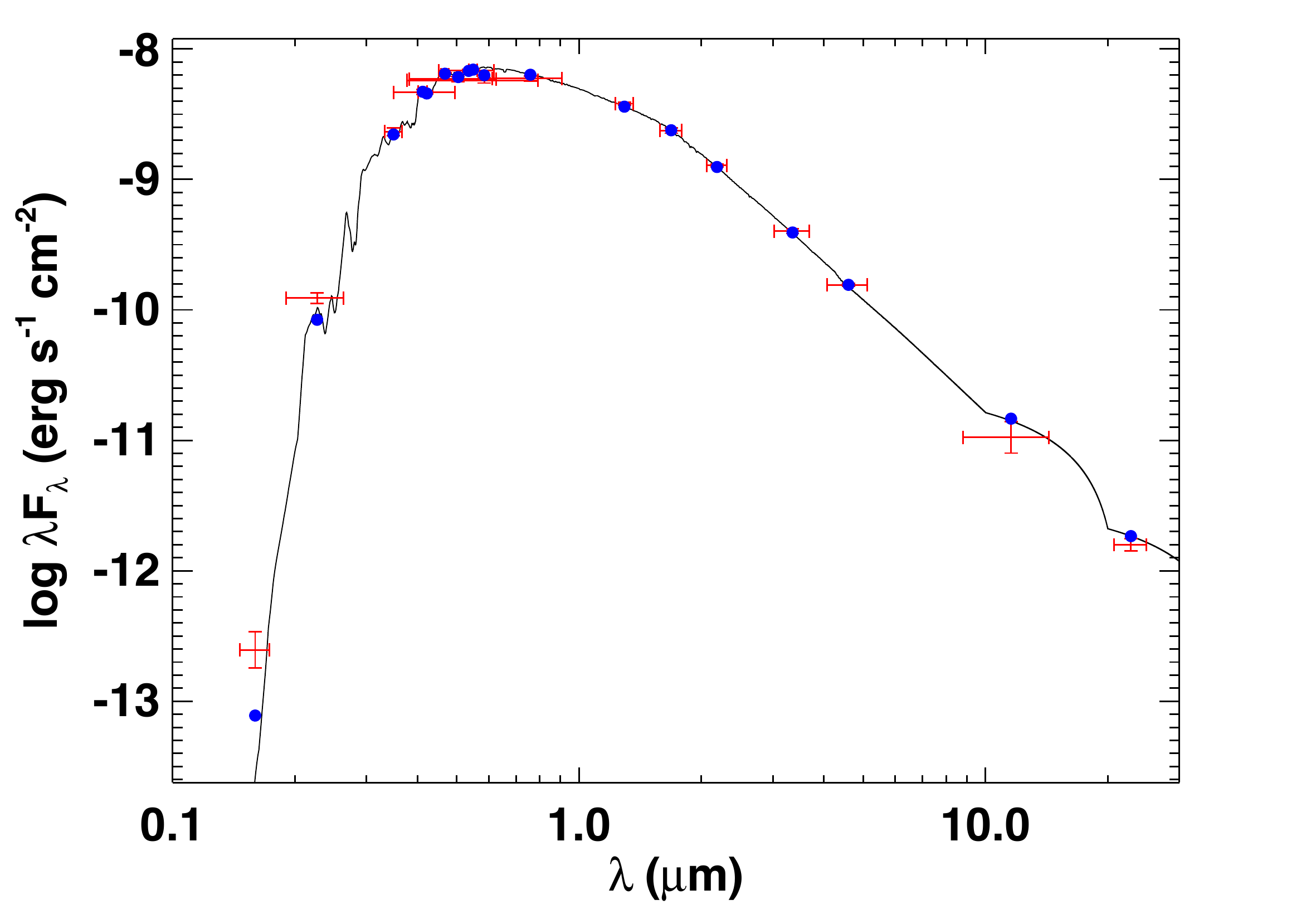}
\caption{The best SED fit for \tar. The Red symbols show the observed photometric measurements, where the horizontal bars represent the effective width of the passband. The Blue points are the predicted integrated fluxes at the corresponding bandpass. The black line represents the best-fit NextGen atmosphere model.}
\label{fig:sed}
\end{figure}

For comparison, we then performed an analysis of the broadband spectral energy distribution (SED) together with the \gaia\ EDR3 parallax \citep{GaiaEDR3} in order to determine an empirical measurement of the stellar radius, following the procedures described in \cite{Stassun2016} and \cite{Stassun2017,Stassun2018}. We gathered the FUV, NUV magnitudes from {\it GALEX} \citep{Morrissey2007}, the $B_{\rm T}$, $V_{\rm T}$ magnitudes from {\it Tycho-2} \citep{hog2000}, the Str\"omgren $u$,$v$,$b$,$y$ magnitudes from \citet{Paunzen2015}, the $J$,$H$,$K_{S}$ magnitudes from {\it 2MASS} Point Source Catalog \citep{Cutri2003,skrutskie2006}, four {\it Wide-field Infrared Survey Explorer} ({\it WISE}) magnitudes \citep{wright2010} and three \gaia\ magnitudes $G$, $G_{\rm BP}$, $G_{\rm RP}$. Together, the available photometry spans the full stellar SED over the wavelength range 0.15--22~$\mu$m (see Figure~\ref{fig:sed}). 

We performed a fit using the Kurucz stellar atmosphere models, with the priors on effective temperature ($T_{\rm eff}$), surface gravity ($\log g$) and metallicity ([Fe/H]) from the spectroscopic analysis. The remaining free parameter is the extinction ($A_V$), which we limited to the maximum permitted for the star's line of sight from the dust maps \citep{Schlegel1998}. The best-fit SED is shown in Figure~\ref{fig:sed} with a reduced $\chi^2 = 1.4$ (excluding the {\it GALEX\/} UV measurements, which indicate mild chromospheric activity; see below) and $A_V = 0.01\pm0.01$. Integrating the model SED gives a bolometric flux at Earth of $F_{\rm bol} = 9.32 \pm 0.11 \times 10^{-9}$ erg~s$^{-1}$~cm$^{-2}$. Taking the $F_{\rm bol}$ and $T_{\rm eff}$ together with the {\it Gaia\/} parallax, we obtained a stellar radius of $R_\ast = 0.972 \pm 0.014\ R_{\odot}$, which agrees with the previous result within $1\sigma$. 

We computed an empirical estimate of the stellar mass from this $R_\ast$ together with the spectroscopic $\log g$, from which we obtained $M_\ast = 1.03 \pm 0.05\ M_{\odot}$. This is consistent with that estimated via the eclipsing-binary based relations of \citet{Torres2010}, which gives $M_\ast =1.04 \pm 0.06\ M_{\odot}$.

Taking all the results above into consideration, we finally adopted the weighted mean values of effective temperature $T_{\rm eff}$, stellar radius $R_{\ast}$ and stellar mass $M_{\ast}$. Combining the expected stellar radius with mass, we found a mean stellar density of $\rm \rho_\ast = 1.58 \pm 0.16$~g~cm$^{-3}$. 

Following \cite{Johnson1987}, we adopted the astrometric values ($\varpi$, $\mu_{\alpha}$, $\mu_{\delta}$) from \gaia\ EDR3 \citep{GaiaEDR3} as well as systemic RV taken from \gaia\ DR2 \citep{Gaia2018}, and computed the three-dimensional Galactic space motion of ($U_{\rm LSR}$, $V_{\rm LSR}$, $W_{\rm LSR}$) = ($-23.10\pm0.19$, $1.53\pm0.06$, $-14.14\pm0.10$) km s$^{-1}$, all of which are relative to the LSR. Building on the kinematic calculation, we then determined the relative probability $P_{\rm thick}/P_{\rm thin}$ of \tar\ to be in the thick and thin disks \citep{Bensby2003,Bensby2014}. We obtained $P_{\rm thick}/P_{\rm thin}=0.01$, indicating a thin-disk origin. We further employed the \code{galpy} package \citep{Bovy2015} to estimate the maximal height $Z_{\rm max}$ of \tar\ above the Galactic plane, along with the ``MWPotential2014'' Galactic potential following \cite{Gan2020a}. We find \tar\ has a $Z_{\rm max}$ of $\sim213$ pc, which agrees with our thin-disk conclusion.

The {\it GALEX} photometry suggests a mild amount of chromospheric activity. Indeed, \cite{Lorenzo2018} reported a spectroscopically measured $\log R'_{\rm HK}=-4.89\pm0.02$, consistent with a mild level of activity. Based on the Yonsei-Yale isochrones, \cite{Spina2018} found that \tar\ has an age of $2.6\pm0.5$~Gyr. We list all final adopted stellar parameter values in Table~\ref{starparam}. 


\begin{table}
    \caption{Stellar parameters of \tar}
    \begin{tabular}{lll}
        \hline\hline
        Parameter       &Value       &Reference \\\hline
        \it{Star ID}                    \\
         TIC                     &320004517   \\
         TOI                     &1055         \\
         HIP                    &96160          \\
         \it{Astrometric properties}\\
         $\alpha\ (J2000)$                    &$\ \ $19:33:08.58 \\
         $\delta\ (J2000)$                    &$-$54:31:56.50    \\
         $\varpi$ (mas)              &$17.609\pm0.016$  &\gaia\ EDR3  \\
         $\mu_{\alpha}$ (mas~yr$^{-1}$)     &$108.32\pm0.01$   &\gaia\ EDR3   \\
         $\mu_{\delta}$ (mas~yr$^{-1}$)     &$-82.71\pm0.01$   &\gaia\ EDR3  \\
         RV\ (km~s$^{-1}$)                          &$-15.8\pm0.2$ &\gaia\ DR2  \\
         \it{Photometric properties}\\
         $\tess$\ (mag)           &$8.089\pm0.006$   &$\rm TIC\ V8^{[1]}$     \\
         $G$\ (mag)           &$8.5265\pm0.0002$   &\gaia\ EDR3   \\
         $G_{\rm BP}$\ (mag)           &$8.843\pm0.001$   &\gaia\ EDR3   \\
         $G_{\rm RP}$\ (mag)           &$8.037\pm0.002$   &\gaia\ EDR3   \\
         $B_{T}$\ (mag)                 &$9.477\pm0.019$   &Tycho-2  \\
         $V_{T}$\ (mag)                 &$8.750\pm0.013$   &Tycho-2  \\
         $J$\ (mag)                    &$7.518\pm0.023$   &2MASS \\
         $H$\ (mag)                    &$7.231\pm0.047$   &2MASS \\
         $K_{S}$\ (mag)                &$7.150\pm0.027$    &2MASS \\
         $W1$ (mag)                   &$7.090\pm0.043$   &{\it WISE} \\
         $W2$ (mag)                   &$7.137\pm0.020$   &{\it WISE} \\
         $W3$ (mag)                   &$7.138\pm0.019$   &{\it WISE} \\
         $W4$ (mag)                   &$7.040\pm0.114$   &{\it WISE} \\
         \it{Derived parameters} \\
         $\log g_{\ast}$ (cgs)       &$4.47\pm0.03$  &This work        \\
         $\rm [Fe/H]$ (dex)  &$-0.07\pm0.09$ &This work
         \\
         Distance (pc)                &$56.79\pm0.06$  &This work     \\
         $U_{\rm LSR}$ (km~s$^{-1}$)       &$-23.10\pm0.19$     &This work\\
         $V_{\rm LSR}$ (km~s$^{-1}$)       &$1.53\pm0.06$     &This work\\
         $W_{\rm LSR}$ (km~s$^{-1}$)       &$-14.14\pm0.10$     &This work\\
         $T_{\rm eff}^{[2]}$ (K)           &$5706\pm110$  &This work       \\
         $M_{\ast}\ (M_{\odot})$ &$1.034\pm0.050$ &This work       \\
         $R_{\ast}\ (R_{\odot})$ &$0.974\pm0.015$ &This work       \\
         $\rm \rho_\ast\ (g~cm^{-3})$ &$1.58\pm0.16$ &This work 
         \\
         $P_{\rm rot}$ (day) &$23.2\pm3.7$ &This work 
         \\
         $A_{V}$ (mag) &$0.01\pm0.01$ &This work \\
         Age (Gyr)     &$2.6\pm0.5$   &\cite{Spina2018}
         \\
         \hline\hline 
    \end{tabular}
    \begin{tablenotes}
    \item[1]  [1]\ \cite{Stassun2017tic,Stassun2019tic}
    \item[2]  [2]\ We take the average values of $T_{\rm eff}$, $M_{\ast}$ and $R_{\ast}$ here (see Section \ref{stellar_characterization}). 
    \end{tablenotes}
    \label{starparam}
\end{table}

\subsection{Stellar Rotation}\label{rotation}
The \tess\ PDCSAP light curve from sector 13 shows a clear variation with a timescale of $\sim 9.5$ d, which implies a relatively high stellar rotation speed. However, this periodic signature is not shown in the corresponding SAP light curve (see Figure \ref{transit_detrend}). Additionally, the subsequent light curve from the extended mission does not have a similar trend. We show below that this $\sim 9.5$ d signal is more likely due to instrumental systematic errors instead of real stellar variability. 

First, we estimated the rotation period $P_{\rm rot}/\sin i = 24.4 \pm 2.6$~d based on the stellar radius $R_{\ast}$ together with the spectroscopically determined rotational velocity $v\sin i = 2.1 \pm 0.2$~km~s$^{-1}$ \citep{Soto2018}. Assuming $\sin i = 1$, this is consistent with the value $P_{\rm rot} = 23.2 \pm 3.7$~d inferred using the empirical activity-rotation relation from \cite{Mamajek2008} according to gyrochronology \citep{Barnes2007,Meibom2009,Curtis2019}.

Furthermore, \cite{McQuillan2014} analyzed the rotation periods of main-sequence stars below 6500 K based on three years of data from the \kepler\ space mission. Our derived rotation period $P_{\rm rot}$ $23.2\pm3.7$ d of \tar\ agrees with the typical value $\sim 20$ d of solar-like stars with a $T_{\rm eff}$ of $\sim 5700$ K (see Figures 4 and 5 in \citealt{McQuillan2014}).

We also investigated the rotational modulation in the WASP accumulated data as archival long term light curve could provide information on stellar rotation features. However, we did not find significant signals likely due to the influence of lunar stray light in the 23-27 d regime. 

\begin{figure}
\centering
\includegraphics[width=0.5\textwidth]{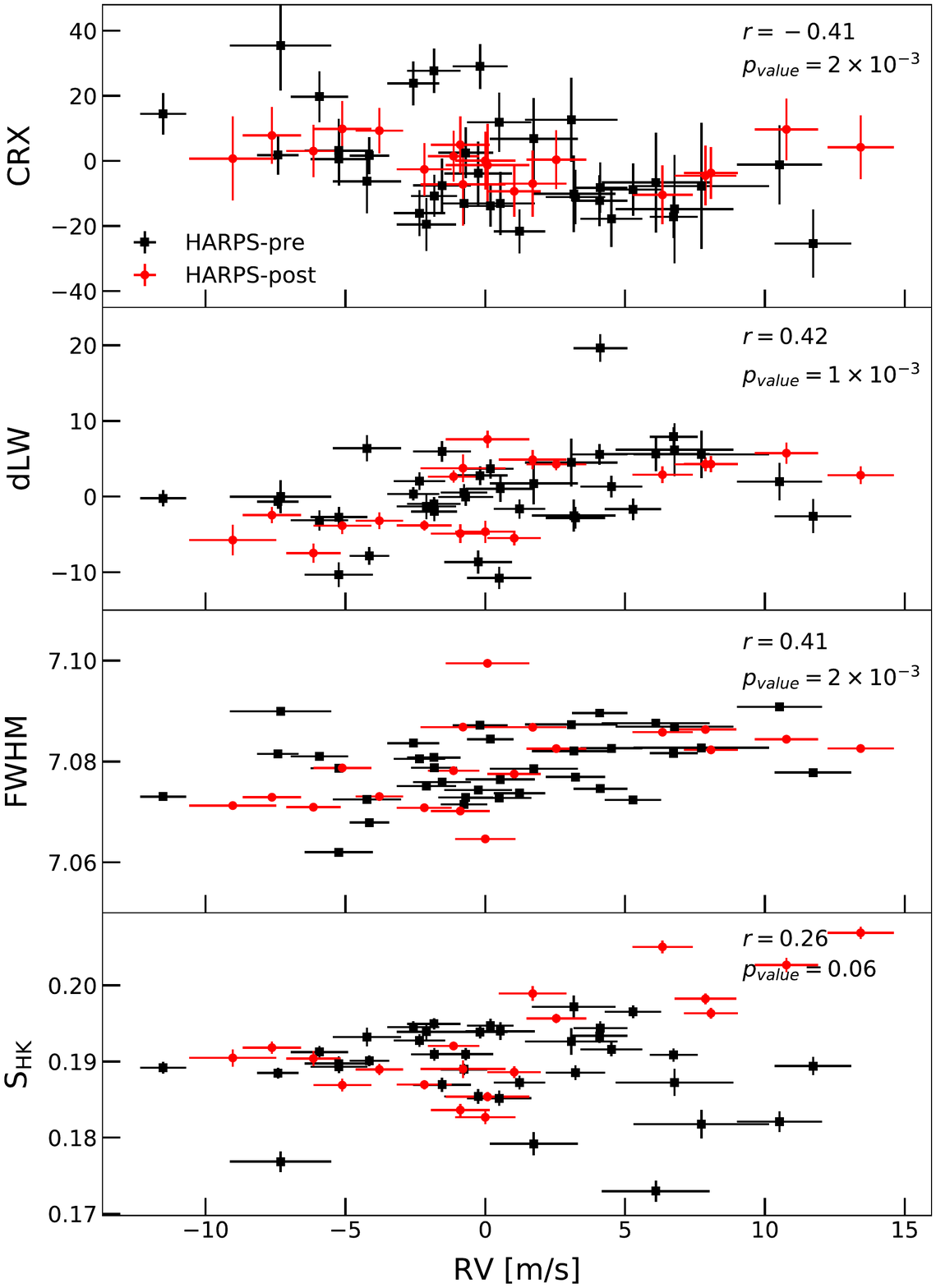}
\caption{Correlations between HARPS RVs and activity indicators (CRX, dLW, FWHM and $S_{\rm HK}$). Different colors represent HARPS pre-upgrade/post-upgrade data. The Pearson's correlation indices and the corresponding p-values are shown on the upper right. In each plot we have subtracted the median value of both RV and the activity indices. The clear correlations indicate that stellar activity has an effect on the Doppler signals (see Section \ref{rotation}).} 
\label{har_ai}
\end{figure}

Finally, we performed a frequency analysis for the HARPS activity indicators (CRX, dLW, bisector span, FWHM and $S_{\rm HK}$). One instrumental effect must be accounted for here: In June 2015, the HARPS optical fibers were replaced as part of a major instrument upgrade, leaving an effective offset between RVs and other line profile-sensitive components measured before and after the upgrade \citep{Lo2015}. For this initial inspection, we calculated the median values for both RVs and each indicator, and subtracted the corresponding offset between pre-upgrade and post-upgrade data. Because of the sparse sampling, we did not see any significant periodic signals, therefore, we do not present the periodograms here. However, we found strong correlations between RVs and CRX, dLW and FWHM ($r = -0.41$, p-value $=2\times10^{-3}$; $r = 0.42$, p-value $= 1\times10^{-3}$; $r = 0.41$, p-value $= 2\times10^{-3}$) and weak correlations between RVs and $S_{\rm HK}$ ($r = 0.26$, p-value $= 0.06$) as shown in Figure \ref{har_ai}, which motivated us to take the stellar activity into consideration in the following RV modeling (see Section \ref{rv}).


\section{Analysis}\label{analysis}
\subsection{Photometric Analysis}\label{transit}
We utilized the \code{juliet} package \citep{juliet} to perform a joint-fit of both space and ground-based light curves. The transit is modeled by \code{batman} \citep{Kreidberg2015}. We applied the dynamic nested sampling approach to determine the posterior probability distribution of the system parameters using the public package \code{dynesty} \citep{Higson2019,Speagle2019}.

We retrieved a list of nearby stars of \tar\ ($G_{\rm rp}=8.037$ mag) within 30$''$ in \gaia\ EDR3 to estimate the flux dilution effect in the ground-based photometries \citep{juliet}. Three faint stars with $G_{\rm rp}>17.4$ mag are found located at $>25''$ away from \tar. As the nearby stars are faint and relatively distant, these stars should make minor contribution to the contaminated flux, which is consistent with the small contamination ratio $A_{D}=0.001$ reported in the \tess\ Input Catalog (TIC) V8 \citep{Stassun2017tic,Stassun2019tic}. Thus we fixed the dilution factors $D_{\rm LCO}$ equal to 1 but considered individual instrument offsets. 

\begin{figure*}
\centering
\includegraphics[width=\textwidth]{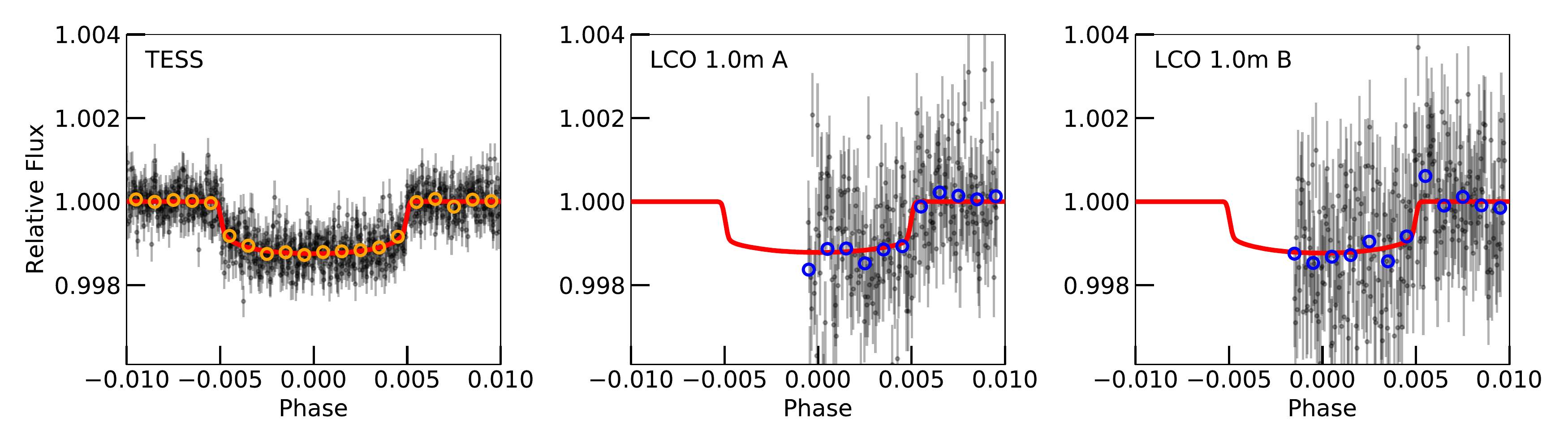}
\caption{Phase-folded transits of \tar b for all available photometric instruments. \tess\ data are presented in the left panel. Two LCO 1m/Sinistro light curves obtained in the Pan-STARRS Y band are shown in the middle and right panels. The orange and blue points represent the binned light curves. The best-fit models are shown as red solid lines.} 
\label{transit_fit}
\end{figure*}

We adopted Gaussian priors for the period $P_{b}$ and mid-transit time $t_{0}$ based on the results from the Box Least Square search. \code{juliet} applies the new parametrizations $r_{1}$ and $r_{2}$ to sample points \citep{Espinoza2018}, for which we set uniform priors between 0 and 1. We adopted a quadratic limb-darkening law for \tess\ photometry and uniformly sampled the coefficients ($q_{1}$ and $q_{2}$, \citealt{Kipping2013}). For ground-based data, we used a linear law instead to parameterize the limb-darkening effect and placed a Gaussian prior on the coefficient, centered at the theoretical estimate derived from the \code{LDTK} package \citep{Husser2013,Parviainen2015} with a $1\sigma$ value of 0.1. We fit a circular orbit for \tar b with a non-informative log-uniform prior set on the stellar density. For each instrument, we included a flux jitter term to account for the white noise. The results of the fit along with the corresponding prior settings are listed in Table \ref{tranpriors}. We present the best-fit models in Figure \ref{transit_fit}.

\begin{table*}
    \centering
    \caption{Model parameters, prior settings and the best-fit values for the \tess\ and ground-based light curves of \tar.}
    \begin{tabular}{lccr}
        \hline\hline
        Parameter       &Best-fit Value       &Prior     &Description\\\hline
        \it{Planetary parameters}\\
        $P_{b}$ (days)   &$17.47128^{+0.00005}_{-0.00005}$  
        &$\mathcal{N}^{[1]}$ ($17.4$\ ,\ $0.1^{2}$)
        &Orbital period of \tar b.\\
        $T_{0,b}$ (BJD-2457000)    &$1661.06295^{+0.00070}_{-0.00071}$ 
        &$\mathcal{N}$ ($1661.1$\ ,\ $0.1^{2}$) 
        &Mid-transit time of \tar b.\\
        $r_{1,b}$    &$0.613^{+0.133}_{-0.127}$ 
        &$\mathcal{U}^{[2]}$ (0\ ,\ 1)
        &Parametrisation for {\it p} and {\it b}.\\
        $r_{2,b}$    &$0.03319^{+0.00069}_{-0.00056}$ 
        &$\mathcal{U}$ (0\ ,\ 1)
        &Parametrisation for {\it p} and {\it b}.\\
        $e_{b}$                     &0  &Fixed  &Orbital eccentricity of \tar b.\\
        $\omega_{b}$ (deg)          &90 &Fixed  &Argument of periapsis of \tar b.\\
        
        \it{\tess\ photometry parameters}\\
        $D_{\rm TESS}$     &$1$ 
        &Fixed      &\tess\ photometric dilution factor.\\
        $M_{\rm TESS}$    &$-0.0000003^{+0.000002}_{-0.000002}$
        &$\mathcal{N}$ (0\ ,\ $0.1^{2}$)      &Mean out-of-transit flux of \tess\ photometry.\\
        $\sigma_{\rm TESS}$ (ppm) &$110^{+5}_{-5}$
        &$\mathcal{J}^{[3]}$ ($10^{-6}$\ ,\ $10^{6}$)      &\tess\ additive photometric jitter term.\\
        $q_{1}$                &$0.32^{+0.18}_{-0.12}$       &$\mathcal{U}$ (0\ ,\ 1)  &Quadratic limb darkening coefficient.\\
        $q_{2}$                &$0.26^{+0.32}_{-0.17}$       &$\mathcal{U}$ (0\ ,\ 1)  &Quadratic limb darkening coefficient.\\
        
        \it{LCOA photometry parameters}\\
        $D_{\rm LCOA}$     &$1$ 
        &Fixed      &LCOA photometric dilution factor.\\
        $M_{\rm LCOA}$    &$-0.0006^{+0.00008}_{-0.00007}$
        &$\mathcal{N}$ (0\ ,\ $0.1^{2}$)      &Mean out-of-transit flux of LCOA photometry.\\
        $\sigma_{\rm LCOA}$ (ppm) &$520^{+147}_{-234}$
        &$\mathcal{J}$ ($0.1$\ ,\ $10^{5}$)      &LCOA additive photometric jitter term.\\
        $q_{\rm LCOA}$                &$0.42^{+0.07}_{-0.09}$       &$\mathcal{N}$ ($0.37$\ ,\ $0.1^{2}$)  &Linear limb darkening coefficient.\\
        
        \it{LCOB photometry parameters}\\
        $D_{\rm LCOB}$     &$1$ 
        &Fixed      &LCOB photometric dilution factor.\\
        $M_{\rm LCOB}$    &$-0.0006^{+0.00008}_{-0.00008}$
        &$\mathcal{N}$ (0\ ,\ $0.1^{2}$)      &Mean out-of-transit flux of LCOB photometry.\\
        $\sigma_{\rm LCOB}$ (ppm) &$817^{+87}_{-85}$
        &$\mathcal{J}$ ($0.1$\ ,\ $10^{5}$)      &LCOB additive photometric jitter term.\\
        $q_{\rm LCOB}$                &$0.35^{+0.09}_{-0.08}$       &$\mathcal{N}$ ($0.37$\ ,\ $0.1^{2}$)  &Linear limb darkening coefficient.\\
        \it{Stellar parameters}\\
        ${\rho}_{\ast}$ ($\rm kg\ m^{-3}$)   &$1514^{+353}_{-535}$
        &$\mathcal{J}$ ($100$\ ,\ $\rm 100^{2}$) &Stellar density.\\
        \hline
        \it{Derived parameters}\\
        $R_{p}/R_{\ast}$ &\multicolumn{2}{c}{$0.03319^{+0.00069}_{-0.00056}$} &Planet radius in units of stellar radii.\\
        $R_{p}$ ($R_{\oplus}$) &\multicolumn{2}{c}{$3.53^{+0.13}_{-0.11}$} &Planet radius.\\
        ${b}$    &\multicolumn{2}{c}{$0.42^{+0.20}_{-0.19}$} &Impact Parameter.\\
        $a/R_{\ast}$    &\multicolumn{2}{c}{$29.0^{+2.1}_{-2.8}$} &Semi-major axis in units of stellar radii.\\
        $a$ (AU)      &\multicolumn{2}{c}{$0.13^{+0.01}_{-0.01}$} &Semi-major axis.\\
        $i$ (deg)    &\multicolumn{2}{c}{$89.17^{+0.40}_{-0.58}$} &Inclination angle.\\
        $T_{\rm eq}^{[4]}$ (K)    &\multicolumn{2}{c}{$748^{+55}_{-40}$} &Equilibrium temperature.\\
        \hline\hline 
    \end{tabular}
    \begin{tablenotes}
    \item[1]  [1]\ $\mathcal{N}$($\mu\ ,\ \sigma^{2}$) means a normal prior with mean $\mu$ and standard deviation $\sigma$. 
    \item[2]  [2]\ $\mathcal{U}$(a\ , \ b) stands for a uniform prior ranging from a to b.
    \item[3]  [3]\ $\mathcal{J}$(a\ , \ b) stands for a Jeffrey's prior ranging from a to b.
    \item[4]  [4]\ We assume there is no heat distribution between the dayside and nightside, and that the albedo is zero.
    \end{tablenotes}
    \label{tranpriors}
\end{table*}

\subsection{RV modeling}\label{rv}
We chose to fit the RVs independently of the transit fit with priors informed by the photometric analysis. We employed the \code{forecaster} package to predict the mass of \tar b \citep{Chen2017}. We obtained $12.0^{+9.0}_{-5.3}\ M_{\oplus}$ based on the probabilistic mass-radius relation, which corresponds to a radial velocity semi-amplitude $K_{b}$ of $\sim 2.8^{+2.2}_{-1.2}$ m/s, assuming a circular orbit. This expected RV signal is beyond the detection capability of \textsc{Minerva} (typical error bar is $\sim 7$ m/s). Hence, we chose to analyze the HARPS-only data set first to avoid the \textsc{Minerva} RVs obscuring the signal, then combine the additional \textsc{Minerva} data and perform a joint-fit. 

\subsubsection{HARPS-only}\label{harps-only}

We fit the HARPS-only RVs independent of the transit modeling with priors coming from the best-fit transit ephemeris. \code{Juliet} utilizes the \code{radvel} algorithm to create the Keplerian model \citep{Fulton2018} for the RV time series data. We compared different RV models based on the Bayesian model log-evidence ($\ln Z$) calculated by the \code{dynesty} package. In general, a model is favoured if $\Delta \ln Z>2$ compared with the other, and strongly supported if $\Delta \ln Z>5$ \citep{Trotta2008}. 

\begin{figure*}
\centering
\includegraphics[width=\textwidth]{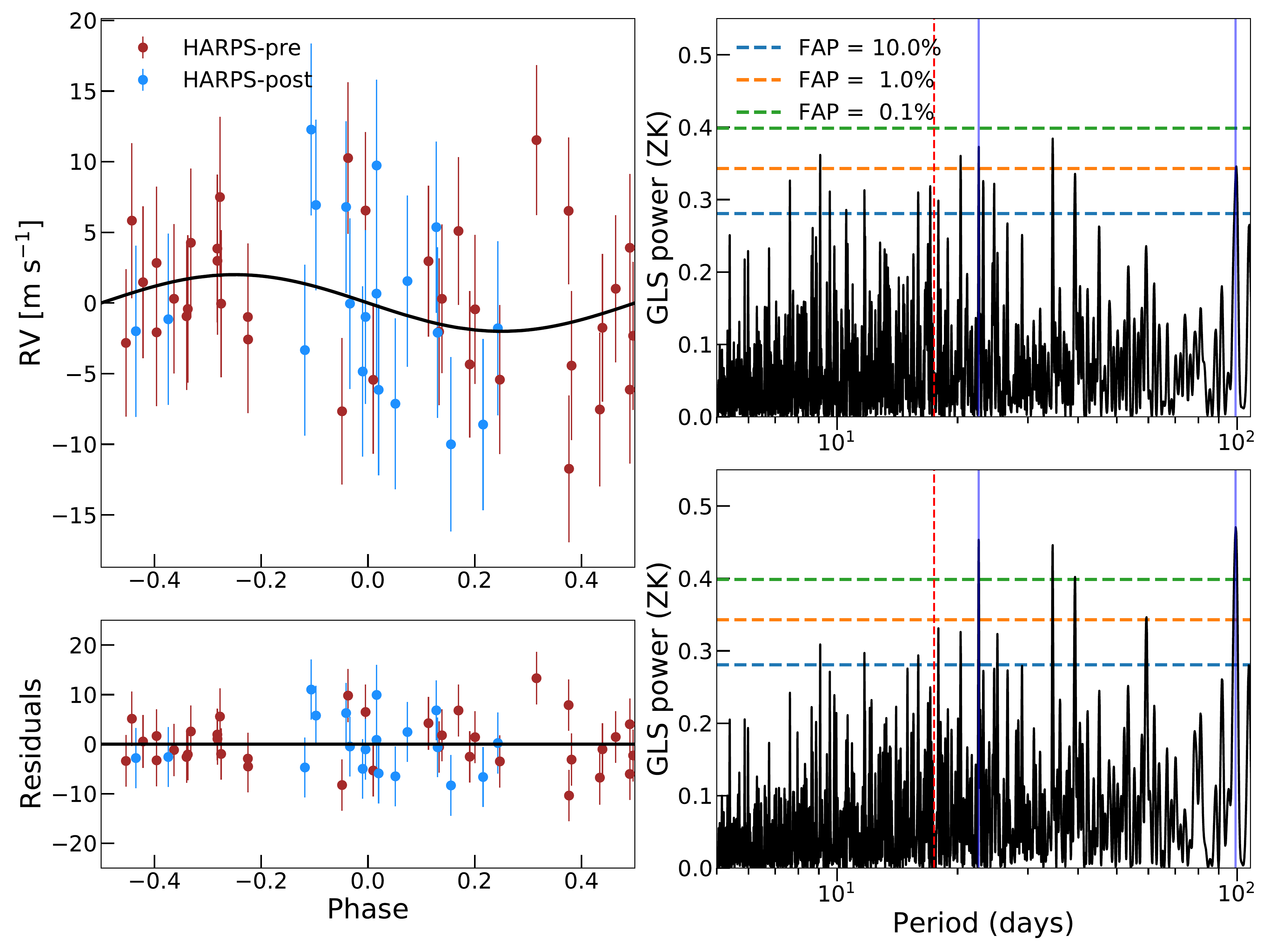}
\caption{{\it Top left panel}: The phase-folded HARPS RVs of \tar. The best-fit base model is shown as a black solid line.  {\it Bottom left panel}: RV residuals after subtracting the best-fit Keplerian model. The error bars are the quadrature sum of the instrument jitter term and the measurement uncertainties for all RVs. {\it Right panels}: The GLS periodograms of the total HARPS RV data (top) and the residuals (bottom) after adjusting for the RV offsets between different instruments using the best-fit values from our base model. The 10\%, 1\% and 0.1\% FAP levels are shown as horizontal dashed lines. The red vertical dashed line represents the period of \tar b ($P_{b}=17.47$~d) derived from the light curve fit. The periodogram of the RV residuals shows up a forest of peaks between 22.6~d and 99.2~d (two blue vertical lines) which may be due to the stellar activity (see Section \ref{rv}).} 
\label{bm}
\end{figure*}

We first performed a simple 1-planet (i.e. \tar b) Keplerian orbit fit with uniform priors on $e\sin \omega$ and $e\cos \omega$. We treated the HARPS-pre and HARPS-post data as from two different instruments and included the RV offset and the RV jitter terms for each set of data. We obtained $e=0.49\pm0.30$, indicating the current RVs are insufficient to constrain the eccentricity. Moreover, compared with a circular orbit model, we found the Bayesian evidence is not significantly stronger for the eccentric model ($\Delta \ln Z=\ln Z_{\rm ecc}-\ln Z_{\rm circ}<1$). Thus we chose to fix the orbital eccentricity to 0 in all our runs and considered this 1-planet circular orbit model as our base model (hereafter BM; 1pl). We further compared the $\ln Z$ of the BM model and a no-planet model (np), and we found a significant improvement ($\Delta \ln Z=\ln Z_{\rm BM}-\ln Z_{\rm np}=13$), supporting the existence of the planet. The BM model gives $K_{b}=2.3^{+1.1}_{-1.0}$ m/s, which leads to a marginal mass measurement of $9.5\pm4.5\ M_{\oplus}$. We show all HARPS RV data along with best-fit model in Figure \ref{bm}. The RV periodogram does not show an obvious planet signal at $\sim 17$ d or any other significant peaks with $\rm FAP<0.1\%$ due to the poor sampling. However, subtracting the best-fit BM model resulted in a forest of peaks between 22.6 d and 99.2 d with $\rm FAP<0.1\%$ in the GLS periodogram of residuals, which may arise from additional planets in the system or from noise that was not accounted for in our model (e.g. stellar activity). 

To investigate the source of these new peaks we identified in the GLS periodogram (see Figure \ref{bm}), we fit a BM+1pl (\tar b + a potential outer planet) model, allowing the period $P_{c}$ vary uniformly between 20 d and 110 d along with a wide uniform prior on the RV semi-amplitude $K_{c}$. However, we did not find any convergences in the $P_{c}-K_{c}$ space, indicating that there is no evidence for the existence of another outer non-transiting planet within the period range. This is also confirmed by the Bayesian model log-evidence, which only shows a negligible improvement compared with the BM model ($\Delta \ln Z=\ln Z_{\rm BM+1pl}-\ln Z_{\rm BM}=2$). 

If stellar activity signals are present in the data, from surface features rotating across the star or from longer-term variations in the net convective blueshift suppression, they could also add peaks to the periodogram. In particular, the phase incoherence of activity signals due to constantly-evolving surface features over the long duration of RV observations will contribute excess power in disordered structures around the rotation period and its harmonics, which unfortunately coincide with the region of period space we wish to search. With all of this in mind, we explored two approaches to deal with the stellar activity effect on the Doppler signals.

First, we modeled the RVs using Gaussian Process regression (BM+GP) with an quasi-periodic kernel formulated by \cite{Foreman2017}:
\begin{equation}
    k_{i,j}(\tau) = \frac{B}{2+C}e^{-\tau/L}\left[{\rm cos}\left(\frac{2\pi\tau}{P_{\rm rot}}\right) + (1+C) \right],
\end{equation}
where $B$ defines the GP covariance amplitude, $C$ is a balance parameter for the periodic and the non-periodic parts, $\tau=|t_{i}-t_{j}|$ is the time-lag between data point $i$ and $j$. $L$ and $P_{\rm rot}$ represent the coherence timescale and the stellar rotational period, respectively. We adopted uninformative, wide log-uniform priors to the GP parameters except for the periodic timescale $P_{\rm rot}$, where we chose a narrow Gaussian prior centering at 23.2 d with $\sigma_{P_{\rm rot}}=4$ d according to our findings in Section~\ref{rotation}. We obtained $K_{b}=4.3\pm1.0$ m/s, which is consistent with our estimate from BM+1pl within $2\sigma$. Although we noticed a significant enhancement of $\ln Z$ ($\Delta \ln Z=\ln Z_{\rm BM+GP}-\ln Z_{\rm BM}=8$), we suspected the GP model might have over-fitted the RV data. The expected stellar rotation timescale ($\sim23$ d) is much smaller than the total RV baseline ($>2800$ d) and the current RV data points are too sparse, which implies the stellar activity signal is not well sampled, making GP not robust in such a challenging case. Comparing with a GP-only model, we also found a $\ln Z$ improvement of the BM+GP model ($\Delta \ln Z=\ln Z_{\rm BM+GP}-\ln Z_{\rm GP}=5$). 

\begin{figure*}
\centering
\includegraphics[width=0.8\textwidth]{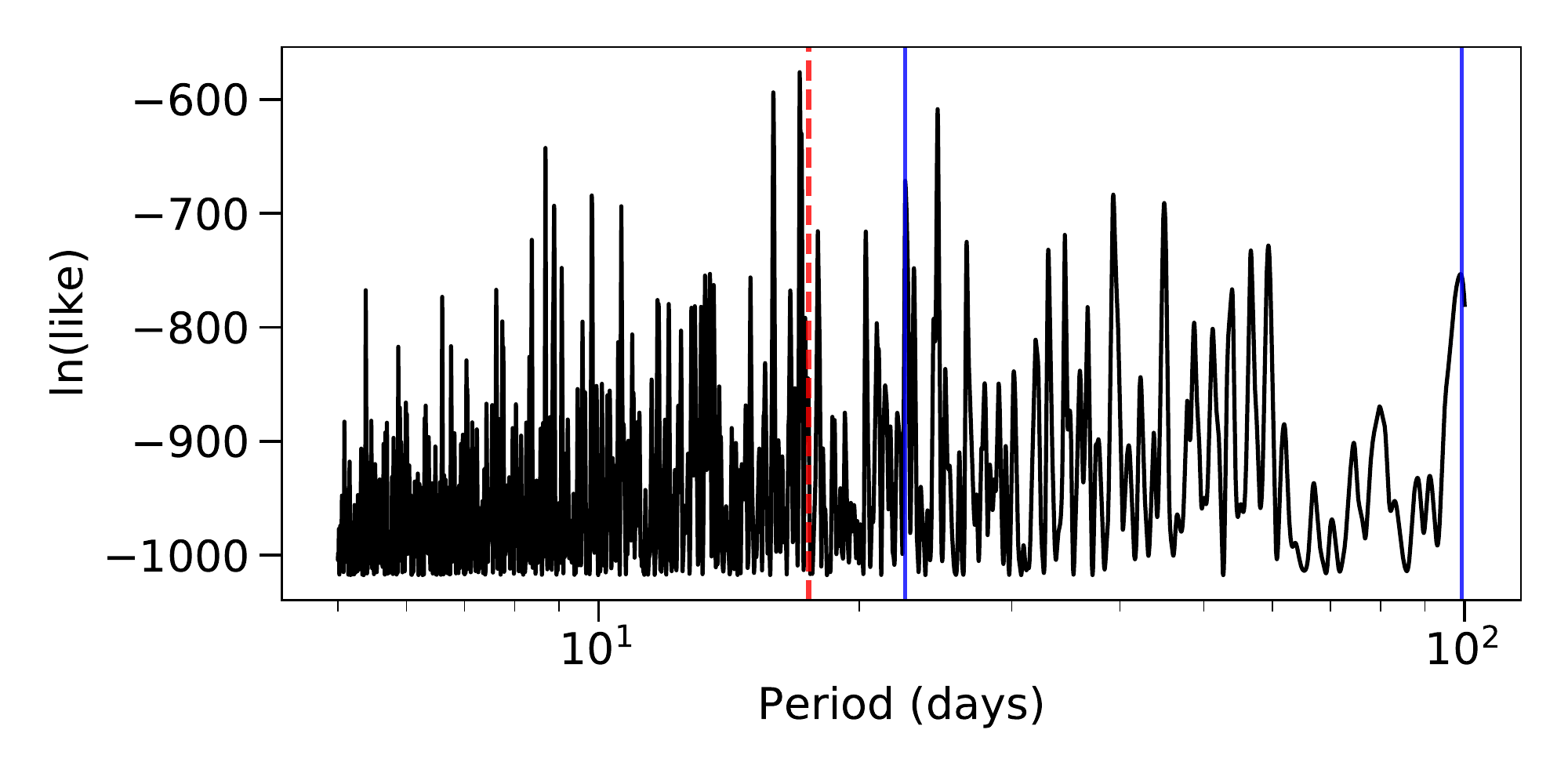}
\caption{Periodogram of the HARPS RVs employing the log-likelihood periodogram method described in the text. Vertical lines mark the period of \tar b (red dashed line) and the bounds of the forest of peaks seen in Figure \ref{bm} (blue solid lines). After accounting for a linear correlation between RV and FWHM and including a quadratic background term, the noise is suppressed to the extent that a 17-day peak can be seen.}
\label{fig:lnlike-periodogram}
\end{figure*}

\begin{figure}
\centering
\includegraphics[width=0.49\textwidth]{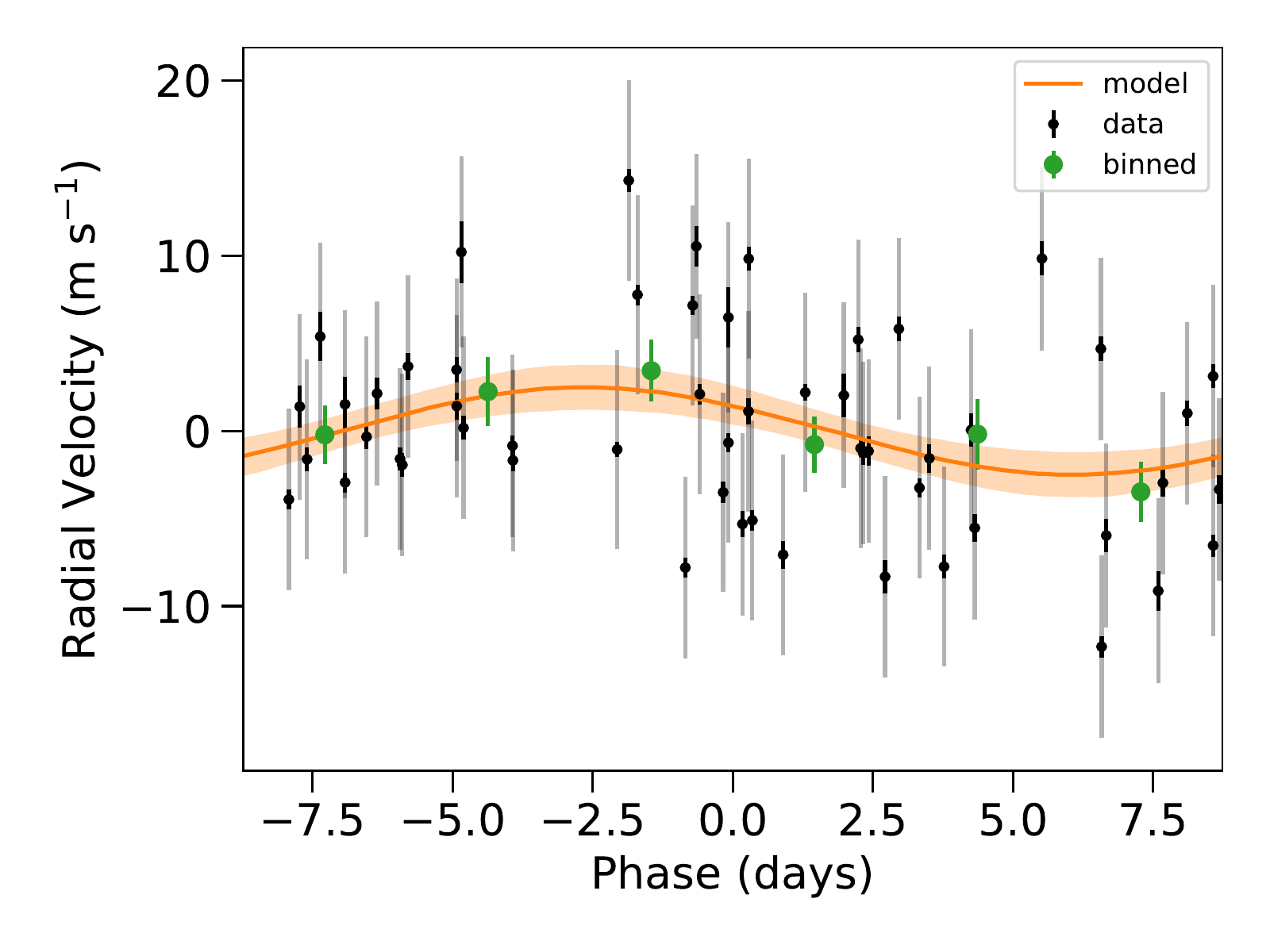}
\caption{Phase-folded radial velocities for \tar b using the BM+FWHM fit. Individual data points and their photon-noise-based uncertainties are shown as black points and error bars, while the grey error bars represent the uncertainties inflated by the best-fit (posterior median) jitter parameters. Green points are the error-weighted means within a series of phase bins. The best-fit BM+FWHM model is shown as a solid orange line, with the shaded region around it marking the model's $1\sigma$ credible interval.}
\label{RV+FWHM}
\end{figure}

We then constructed a simple and fast-to-compute model by involving the activity indicators into the analysis (BM+FWHM). We first tested this option by checking to see whether accounting for activity indicators via linear correlations in the RV analysis would reduce the RV noise to the point that the planet signal could be seen in the periodogram. We adopted a linear relationship between RV and the HARPS FWHM, which was previously shown to correlate significantly with RV (Figure \ref{har_ai}). We also added in a quadratic trend component to account for long-term changes in the RVs due to either the evolving magnetic activity cycle or undetected long-period companions. With this more advanced model, we constructed a \textit{log-likelihood periodogram}. This approach is a simple and efficient way of searching frequency space while taking certain systematic noise sources into account, and follows the methodology behind the Systematics-Insensitive Periodogram introduced for \kepler\ transit searches by \citet{Angus2016}. It is analogous to the Bayesian Generalized Lomb-Scargle periodogram introduced by \citet{Mortier2015}, but contains additional noise terms. Specifically, in our application the model prediction is that for any time $t_n$ with corresponding FWHM measurement $w_n$, the RV $y_n$ is given by:
\begin{equation}
\begin{split}
    y_n = &a x_n^2 + b x_n + c_n + d w_n + K\sin\Big(\frac{2\pi}{P} t_n\Big) +\\
    &H\cos\Big(\frac{2\pi}{P} t_n\Big) + \mathrm{noise},
\end{split}
\end{equation}
where $x_n$ is a normalized relative time:
\begin{equation}
    x_n = \frac{t_n - \langle \mathbf{t} \rangle}{\mathrm{max}(\mathbf{t}) - \mathrm{min}(\mathbf{t})},
\end{equation}
the baseline term $c_n$ is comprised of two possible values based on whether observation $n$ was taken before or after HARPS upgrade time $t_{\rm upgrade}$:
\begin{equation}
    c_n = c_1 \delta_n + c_2 \neg\delta_n,
\end{equation}
with $\delta_n$ set to 1 for pre-upgrade data and 0 for post-upgrade data. 
The variables ($a, b, c_1, c_2, d, K, H, P$) are unknowns to be constrained from the data. 
At any given value of the orbital period $P$, this model is entirely linear, so that the vector of predicted RVs $\mathbf{y}$ can be calculated as a product of a design matrix
\begin{equation}
    \mathbf{A_P} = 
    \begin{pmatrix}
        x_0^2 & x_0 & \delta_0 & \neg\delta_0 & w_0 &  \sin\Big(\frac{2\pi}{P} t_0\Big) &  \cos\Big(\frac{2\pi}{P} t_0\Big)\\
        \vdots & \vdots & \vdots & \vdots & \vdots & \vdots & \vdots \\
        x_n^2 & x_n & \delta_n & \neg\delta_n & w_n &  \sin\Big(\frac{2\pi}{P} t_n\Big) &  \cos\Big(\frac{2\pi}{P} t_n\Big)
    \end{pmatrix}
\end{equation}
and a variable vector
\begin{equation}
    \mathbf{\Theta} = [a, b, c_1, c_2, d, K, H]^T.
\end{equation}
At the period $P_{b}$, then, the optimal parameters $\mathbf{\Theta^*_P}$ can be analytically determined as the following:
\begin{equation}
    \mathbf{\Theta^*_P} = (\mathbf{A_P}^\mathsf{T} \mathbf{C}^{-1} \mathbf{A_P})^{-1} \mathbf{A_P}^\mathsf{T} \mathbf{C}^{-1} \mathbf{y},
\end{equation}
where $\mathbf{C}^{-1}$ is the covariance matrix for the data, here assumed to be diagonal. 
We stepped through a log-uniform grid of periods between 1 and 1000 days and determine the maximum likelihood for each period (neglecting a constant term):
\begin{equation}
    \mathrm{ln}\mathcal{L}^*_P \sim -\frac{1}{2} (\mathbf{y} - \mathbf{A_P}\mathbf{\Theta^*_P})^\mathsf{T} \mathbf{C}^{-1} (\mathbf{y} - \mathbf{A_P}\mathbf{\Theta^*_P}).
\end{equation}
The resulting log-likelihood periodogram shares the fundamental assumption of a circular orbit but is otherwise more robust to stellar activity and long-period trends than a traditional Lomb-Scargle periodogram \citep{Lomb1976,Scargle1982}. Due to the linearity of the model and the resulting ability to find the optimal parameters analytically, the likelihood may be maximized quickly and with a guarantee of convexity. Therefore, while more advanced tools exist to incorporate Bayesian priors \citep{Olspert2018} or non-parametric correlated noise \citep{Feng2017} along with trends in the data, this method is relatively simple to implement, flexible, and fast, making it a practical solution for RV planet searches carried out in the presence of non-negligible stellar noise. The log-likelihood periodogram does show a strong peak at 17 days (Figure \ref{fig:lnlike-periodogram}). We show the resulting fit in Figure \ref{RV+FWHM}. While this supports the presence of a sinusoidal, potentially Keplerian signal in the data, we note that this conclusion depends on the noise model adopted. The 17-day peak is the highest in the log-likelihood periodogram, but a forest of other strong peaks remain. In brief, the above analysis shows that the RV data do support the detection of a 17-day planet, but the signal is not sufficiently strong or robust to changes in the noise model to confidently claim detection on the grounds of RVs alone. 

The results of the log-likelihood periodogram experiment motivated us to include an RV-FWHM correlation term in the RV model (BM+FWHM). 
We implemented this model fit using \code{pymc3} and the \code{exoplanet} package \citep{exoplanet:pymc3, exoplanet:exoplanet}. 
The parameterization and priors adopted were identical those used in the \code{juliet} analysis, with the addition of two free parameters: the slope of the linear correlation between RV and FWHM, $S_{\rm FWHM}$, and an offset of the post-HARPS-upgrade FWHM measurements with respect to the pre-upgrade FWHMs, $\Delta_{\rm FWHM}$. Both of these parameters received broad and uninformative Gaussian priors. 

We regard this BM+FWHM model as our final best model because (1) The BM+FWHM model yields constraints on the planet parameters that are in full agreement with the BM model (Figure \ref{RV+FWHM}, Table \ref{rvpriors}); (2) The BM+FWHM takes stellar activity into consideration while the BM model does not. Though including the FWHM term reduces the white-noise jitter slightly, it is an insignificant reduction, which suggests that the source of the excess noise in the HARPS RV measurements is not sufficiently captured by a FWHM correlation. More RV observations taken with denser sampling to preserve coherency of the stellar activity signal may be necessary to improve the noise model.

\subsubsection{Including the \textsc{Minerva} data}
We finally re-ran a Keplerian fit after including the \textsc{Minerva}-Australis RVs. As \tar\ is monitored by four telescopes (\textsc{Minerva} 1, 3, 4 and 5; denoted as MA, MB, MC and MD) in the array, we treated each data set separately and fit individual offsets and jitters, but kept the same prior settings for other parameters as in Section \ref{harps-only}. We obtained $K_{b}=2.2^{+1.0}_{-0.9}$~m/s, which corresponds to a mass of $9.2^{+4.1}_{-3.7}\ M_{\oplus}$ with a $3\sigma$ upper limit of $21.7\ M_{\oplus}$, estimated using the 99.7\% value of $K_{b}$ in the posterior distribution. This measurement is consistent with the estimate from the HARPS-only data analysis within $1\sigma$. We list our final results in Table \ref{rvpriors} and show the best-fit model of all data in Figure \ref{allrv}. As the \textsc{Minerva} data were noisy with low cadence, we did not attempt to use them to fit the stellar activity.

\begin{table*}
    \centering
    \caption{Parameters, prior settings and the best-fit values for the \tar\ system of three models}
    \begin{tabular}{lcccc}
        \hline\hline
        Parameter       &Priors       &BM for HARPS   &\textbf{BM+FWHM for HARPS}   &BM for HARPS+\textsc{Minerva}\\\hline
        \it{Planetary parameters}\\
        $P_{b}$ (days)   &$\mathcal{N}$ ($17.47128$\ ,\ $0.00005^{2}$) &$17.47129^{+0.00004}_{-0.00004}$ 
        &$17.4712750^{+0.0000097}_{-0.0000153}$
        &$17.47128^{+0.00005}_{-0.00004}$ \\
        $T_{0,b}$ (BJD)   &$\mathcal{N}$ ($2458661.0628$\ ,\ $0.0007^{2}$) &$2458661.06279^{+0.00050}_{-0.00051}$
        &$2458661.06279^{+0.00071}_{-0.00069}$
        &$2458661.06279^{+0.00054}_{-0.00054}$\\
        $e_{b}$                     &Fixed  &0 &0 &0\\
        $\omega_{b}$ (deg)          &Fixed &90 &90 &90\\
        \it{RV offset}\\
        $\rm \mu_{HARPS_{pre}}$ ($\rm m\ s^{-1}$) &$\mathcal{U}$ ($-10$\ ,\ $10$) &$0.64^{+0.89}_{-0.88}$ &$0.9^{+0.9}_{-0.9}$ &$0.65^{+0.90}_{-0.90}$\\
        $\rm \mu_{HARPS_{post}}$ ($\rm m\ s^{-1}$) &$\mathcal{U}$ ($-10$\ ,\ $10$) &$1.28^{+1.33}_{-1.35}$ &$0.7^{+6.5}_{-5.6}$ &$1.30^{+1.32}_{-1.33}$\\
        $\rm \mu_{MA}$ ($\rm m\ s^{-1}$) &$\mathcal{U}$ ($-20$\ ,\ $20$) &- &- &$6.51^{+5.31}_{-5.65}$\\
        $\rm \mu_{MB}$ ($\rm m\ s^{-1}$) &$\mathcal{U}$ ($-20$\ ,\ $20$) &- &- &$0.95^{+4.77}_{-4.88}$\\
        $\rm \mu_{MC}$ ($\rm m\ s^{-1}$) &$\mathcal{U}$ ($-20$\ ,\ $20$) &- &- &$-5.98^{+4.60}_{-4.60}$\\
        $\rm \mu_{MD}$ ($\rm m\ s^{-1}$) &$\mathcal{U}$ ($-20$\ ,\ $20$) &- &- &$-2.17^{+4.83}_{-5.20}$\\
        \it{RV noise}\\
        $\rm \sigma_{HARPS_{pre}}$ ($\rm m\ s^{-1}$) &$\mathcal{U}$ ($0$\ ,\ $10$) &$5.32^{+0.75}_{-0.63}$ &$5.17^{+0.72}_{-0.57}$ &$5.30^{+0.75}_{-0.60}$\\
        $\rm \sigma_{HARPS_{post}}$ ($\rm m\ s^{-1}$) &$\mathcal{U}$ ($0$\ ,\ $10$) &$5.86^{+1.18}_{-0.89}$ &$5.65^{+1.12}_{-0.85}$ &$5.88^{+1.14}_{-0.90}$ \\
        $\rm \sigma_{MA}$ ($\rm m\ s^{-1}$) &$\mathcal{U}$ ($0$\ ,\ $20$) &- &- &$9.97^{+5.33}_{-4.81}$ \\
        $\rm \sigma_{MB}$ ($\rm m\ s^{-1}$) &$\mathcal{U}$ ($0$\ ,\ $20$) &- &- &$12.75^{+3.63}_{-3.45}$ \\
        $\rm \sigma_{MC}$ ($\rm m\ s^{-1}$) &$\mathcal{U}$ ($0$\ ,\ $20$) &- &- &$16.85^{+1.88}_{-2.42}$ \\
        $\rm \sigma_{MD}$ ($\rm m\ s^{-1}$) &$\mathcal{U}$ ($0$\ ,\ $20$) &- &- &$10.12^{+4.69}_{-3.98}$ \\
        \it{Stellar activity}\\
        $S_{\rm FWHM}$ &$\mathcal{N}$ ($0$\ ,\ $10$) &- &$0.4^{+4.9}_{-4.1}$ &- \\
        $\Delta_{\rm FWHM}$ &$\mathcal{N}$ ($0$\ ,\ $5$) &- &$0.1^{+2.4}_{-2.3}$ &- \\
        \it{RV semi-amplitude}\\
        $K_{b}$ ($\rm m\ s^{-1}$)   &$\mathcal{U}$ ($0$\ ,\ $10$)
        &$2.3^{+1.1}_{-1.0}$ &$2.7^{+1.3}_{-1.3}$ &$2.2^{+1.0}_{-0.9}$\\
        \hline
        \it{Derived parameters}  \\
        $M_{p}^{[1]}$ ($M_{\oplus}$) & &$9.6^{+4.5}_{-4.2}$ &$11.2^{+5.4}_{-5.4}$ &$9.1^{+4.2}_{-3.7}$\\
        $\rho_{p}$ ($\rm g\ cm^{-3}$) & &$1.2^{+0.8}_{-0.6}$ &$1.4^{+0.9}_{-0.8}$ &$1.2^{+0.6}_{-0.8}$\\
        \hline\hline 
    \end{tabular}
    \begin{tablenotes}
    \item[1]  [1]\ This is not a statistically significant measurement. $3\sigma$ mass upper-limit is $27.4\ M_{\oplus}$. 
    \end{tablenotes}
    \label{rvpriors}
\end{table*}

\section{Solar analogs with planets}\label{solar_analogs}
A unique aspect of solar twin planet host stars is the ability to resolve their photospheric abundances at very high precision when compared to the Sun. 
The relationship between a star's composition and the nature of its planetary system is an open question \citep{Hinkel2018,Clark2020}, and solar twins present a promising avenue of investigation.

The Sun has been previously found to have a depletion in refractory elements compared to the volatile elements when contrasted with the refractory-to-volatile content of most nearby solar analogs \citep{Melendez2009, Ramirez2009}. This phenomenon is found by looking at the correlation between the abundance ratio $\rm [X/H]$ (or sometimes $\rm [X/Fe]$) and condensation temperature $T_{c}$ across multiple elements $\mathrm{X}$. For the majority of stars surveyed, solar-normalized photospheric abundance $\rm [X/H]$ correlates positively with $T_{c}$, indicating that more refractory (higher $T_{c}$) elements are over-represented in these stars compared to the Sun. It has been suggested that the Sun's relative depletion in refractories can be attributed to terrestrial planet formation, with refractory materials in the Solar protoplanetary disk being preferentially ``locked up'' in planetesimals before the disk material was accreted onto the Sun \citep{Melendez2009}. Indeed, later work from \cite{Chambers2010} demonstrated that the difference would disappear if adding 4 $M_{\oplus}$ of Earth-like material to the solar convection zone. Additionally, alternative theories including planet ingestion \citep{Ramirez2011,Spina2015,Oh2018,Church2020} and galactic chemical evolution (GCE; \citealt{Adibekyan2014,Nissen2015,Spina2016}), might be able to explain this phenomenon. However, \cite{Bedell2018} confirmed that the depletion pattern still exists even if the GCE effect has been corrected. More recently, \cite{Booth2020} proposed that the gap opened during the giant planet formation may limit dust accretion by the host star from the disk area exterior to the forming giant planet, which may also result in the depletion in the star like our Sun.

\begin{figure}
\centering
\includegraphics[width=0.49\textwidth]{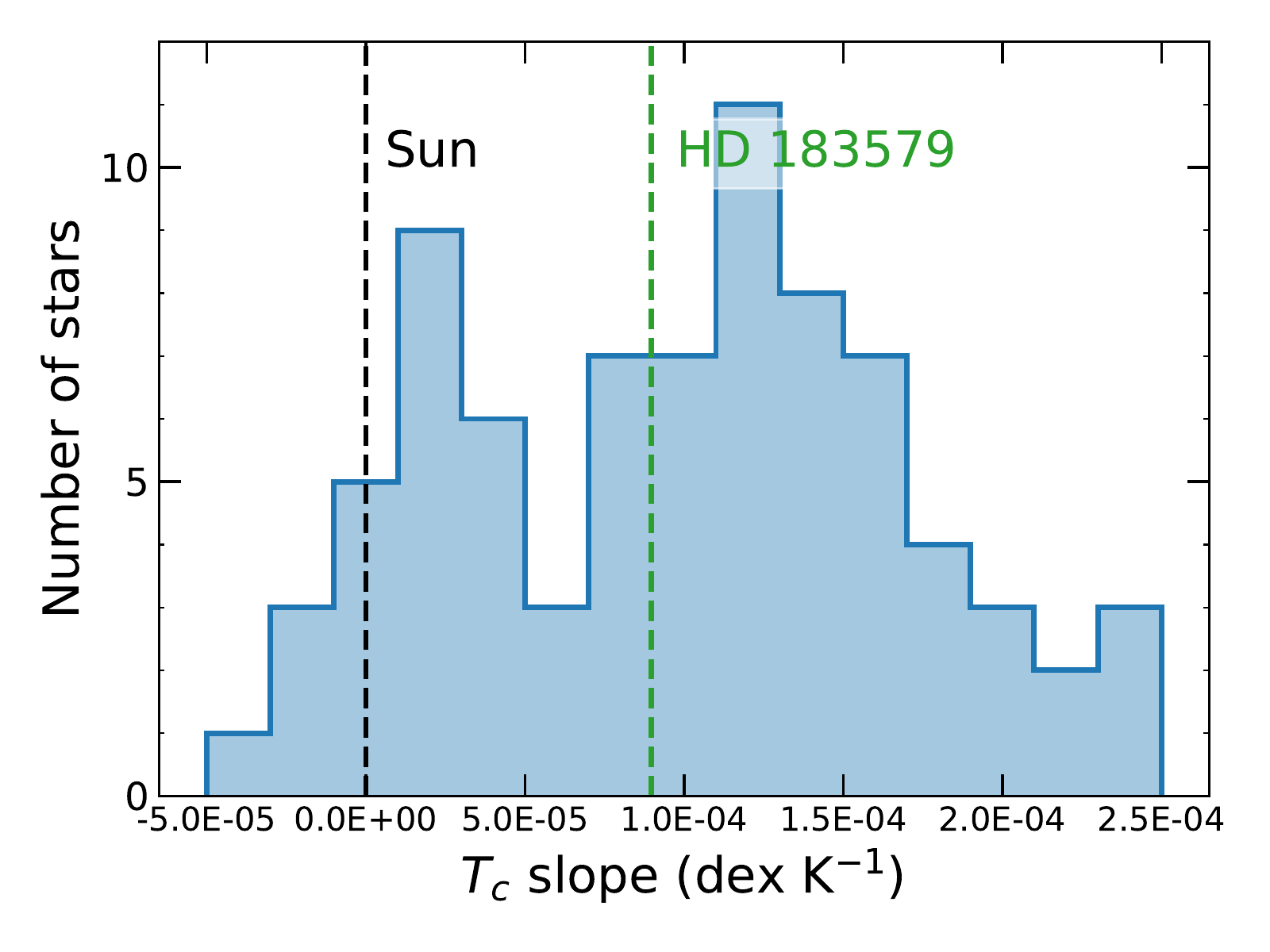}
\caption{Histogram of abundance--condensation temperature ($T_{c}$) trends observed in a 79-star sample of solar twins studied by \citet{Bedell2018}. \tar\ is a member of this sample and its abundance pattern appears typical among solar twins. The abundance patterns for both \tar\ and the general solar twin sample have been corrected for galactic chemical evolution effects as described in \citet{Bedell2018}.}
\label{fig:tc}
\end{figure}

While the exact cause of the Sun's atypical abundance pattern is still unclear, it is informative to look at \tar\ as an example of a solar twin with a markedly different planetary system to the Sun's. 
Many of the HARPS spectra used in this analysis were previously used to analyze the spectroscopic properties and abundances of \tar\ at high precision using a line-by-line differential equivalent width technique \citep{Spina2018, Bedell2018}. 
The abundances of 30 elements were used to examine the behavior of abundance with $T_{c}$ for 79 solar twins, including \tar, in \citet{Bedell2018}. 
Using the galactic chemical evolution-corrected abundance-$T_{c}$ relations measured in that work, we show that \tar\ is a typical Sun-like star, without a statistically significant refractory element depletion (Figure \ref{fig:tc}). 

\begin{figure}
\centering
\includegraphics[width=0.49\textwidth]{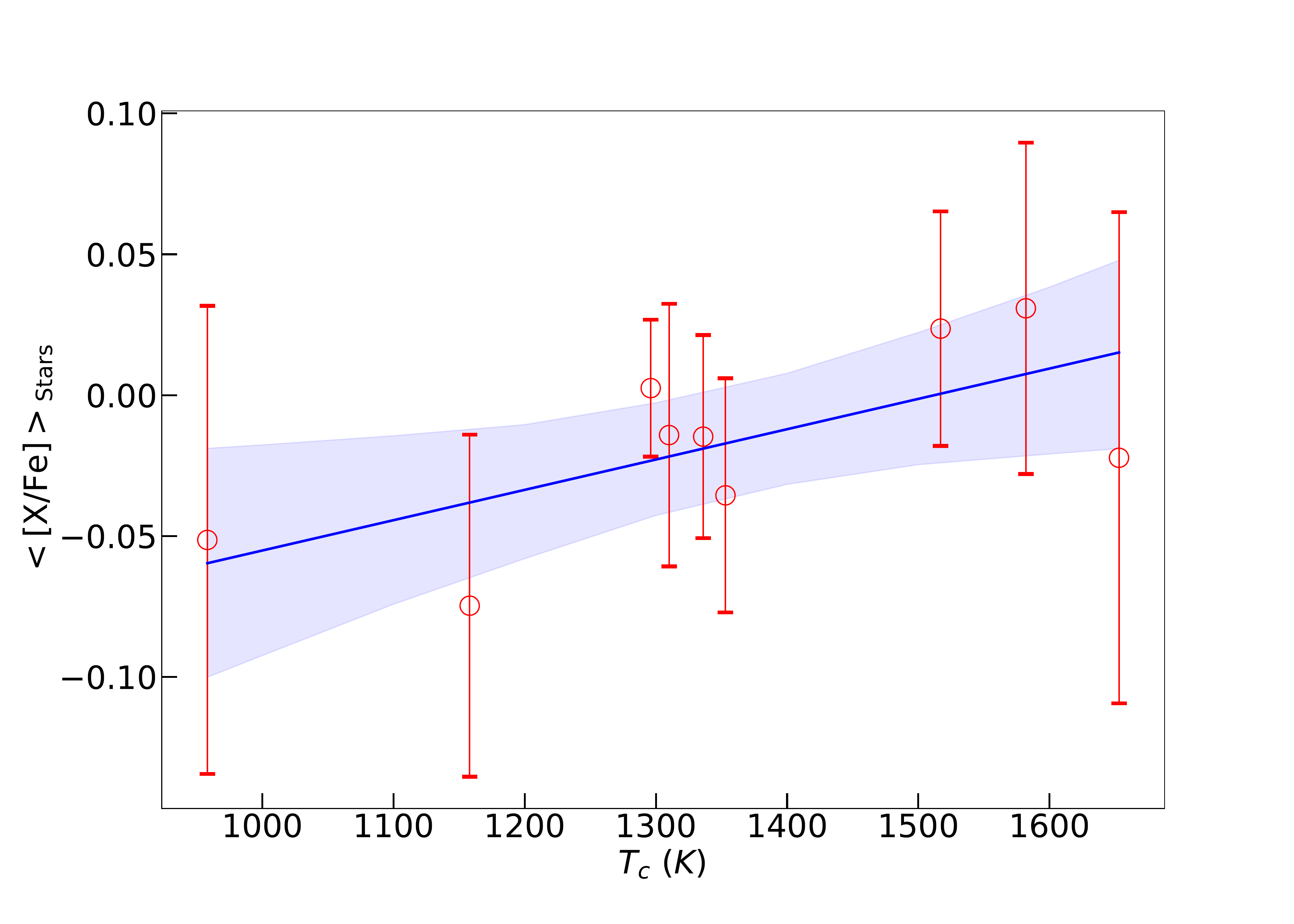}
\caption{The mean abundance of the 36 CKS solar twin planet sample. The error bars represent the standard deviations of each elemental abundance. The blue solid line is the linear fit to the $T_{c}$ trend. The shaded region represents the $1\sigma$ confidence interval.}
\label{fig:trend}
\end{figure}

All of these motivate us to examine if the majority of solar analogs hosting rocky planets (or planets with rocky cores, such as mini-Neptunes) show similar depletion as our Sun, or significant depletion. We used the homogeneous California-Kepler Survey (CKS) catalog to build our planet sample \citep{Petigura2017,Johnson2017}. There are a total of 1305 CKS spectra of ``Kepler Objects of Interest'' (KOIs) that hosting 2025 planet candidates, which precisely measure the stellar properties. We retrieved the publicly available abundances derived by \cite{Brewer2016} and \cite{Brewer2018}, which achieved a typical internal abundance precision at $\sim0.04$ dex level. We first threw out targets flagged as false positives or without dispositions in the catalog, leaving the stars with at least one or more confirmed planets/planet candidates. We then included stars with: (1) $\rm 5680\ K<T_{\rm eff}<5880\ K$; (2) $\rm \sigma_{T_{\rm eff}}<70\ K$; (3) $4.3\ {\rm dex}<\log g<4.5\ {\rm dex}$; (4) $\sigma_{\log g}<0.1\ {\rm dex}$; (5) $-0.1\ {\rm dex}<{\rm [Fe/H]}<0.1\ {\rm dex}$; (6) $\rm \sigma_{[Fe/H]}<0.05\ {\rm dex}$. We found 39 planet-host solar analogs. Since CKS only has a few known giant planet hosts and those systems may bias our comparison as the giant planet formation is also suspected to result in the depletion phenomenon \citep{Booth2020}, we further removed 3 systems with at least one planet with radius larger than 8$R_{\oplus}$ (KOI 1, KOI 372 and KOI 1089), thus our final sample contains 36 stars. We computed the mean abundance of each element (Na, Mg, Al, Si, Ca, Ti, Cr, Mn, Ni), and performed a least-square fit to the $\rm [X/Fe]$ as a function of the condensation temperature $T_{c}$\footnote{We take the 50\% condensation temperatures from \cite{Lodders2003}.}. The result is presented in Figure \ref{fig:trend}. We tentatively found that solar analogs with rocky planets/rocky cores (i.e., mini-Neptunes) do not show a similar depletion as our Sun. However, this preliminary result is limited by the methodology used to derive the chemical abundance. With the current small number of precisely characterized solar twins with known planets, we cannot draw any conclusions. 

\section{Discussion}\label{discussion}

\begin{figure}
\centering
\includegraphics[width=0.49\textwidth]{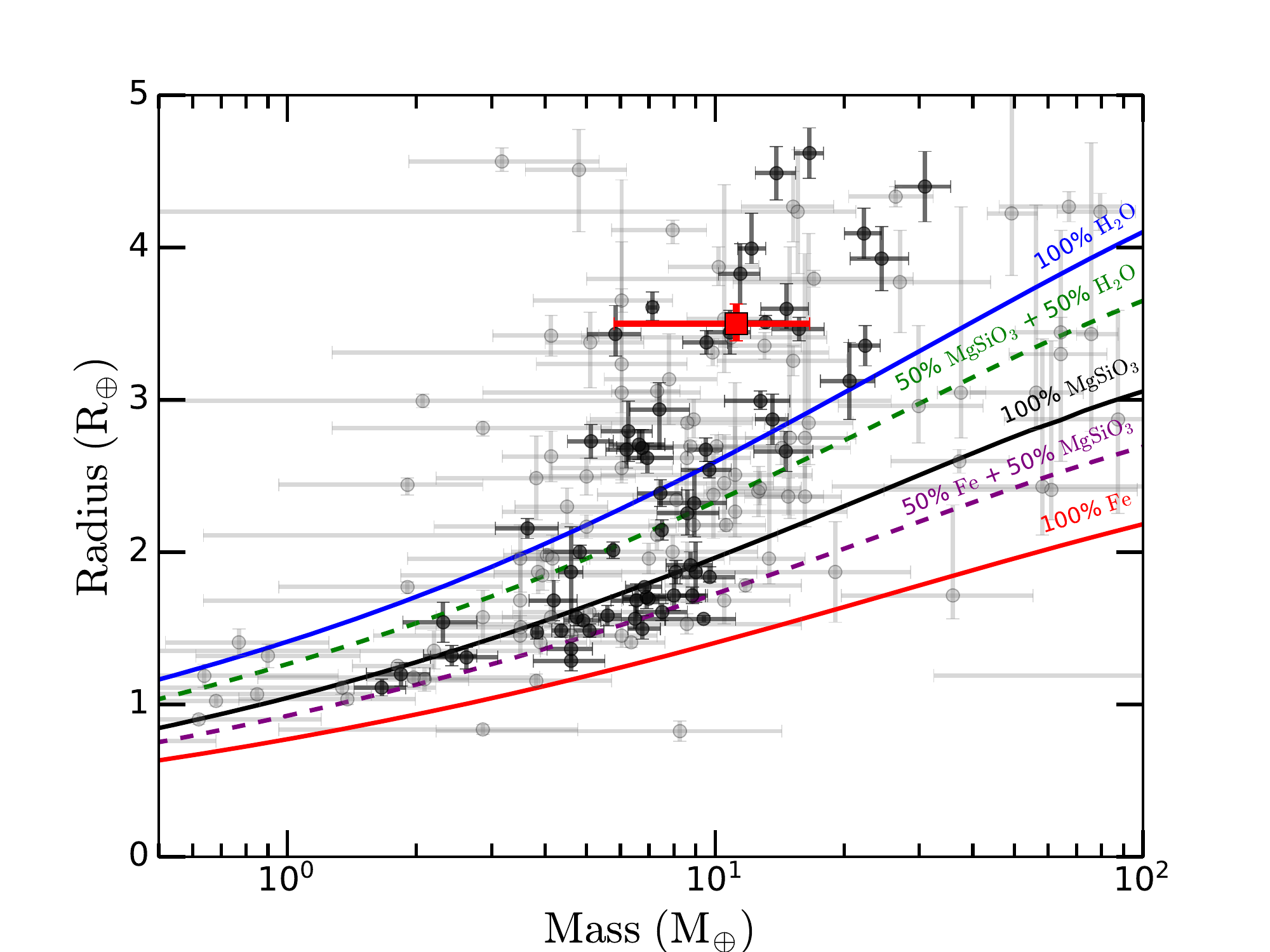}
\caption{The mass radius diagram in Earth units. \tar b is marked as a red square. The confirmed planets with well measured radius and mass are shown as black points (uncertainty smaller than 20
\%) while grey points represent the planet with poor constraint (data are retrieved from NASA Exoplanet Archive; \citealt{Akeson2013}). The colored lines are the theoretical M-R models for different planetary compositions, taken from \citet{Zeng2013}.} 
\label{mr}
\end{figure}

\subsection{Atmospheric characterization of \tar b}
Although thousands of large sub-Neptunes ($2.75\ R_{\oplus}<R_{p}<4\ R_{\oplus}$) have been detected up to now, only $\sim 40$ of them are orbiting around bright stars ($K<10$~mag). With $K=7.15$~mag, \tar b is hosted by the fifth brightest star among them (HD 21749b, \citealt{Trifonov2019,Dragomir2019,Gan2020b}; GJ 436b, \citealt{Butler2004,Knutson2011}; HD 95338b, \citealt{Diaz2020}; and HD 3167c, \citealt{Vanderburg2016HD3167,Livingston2018K2VP}), making it an excellent target for future atmospheric characterization using the upcoming {\it James Webb Space Telescope} (\jwst, \citealt{Gardner2006}) and {\it Extremely Large Telescope} (\elt, \citealt{Gilmozzi2007,Zeeuw2014}). 

Following the approach in \cite{Gillon2016}, we estimated the signal amplitude of \tar b in the transit transmission spectroscopy:
\begin{equation}
    {\rm Amp} = \frac{2R_{p}h_{\rm eff}}{R_{\ast}^{2}},
\end{equation}
where $R_{p}$ and $R_{\ast}$ are the planet and stellar radius, and $h_{\rm eff}=7kT/\mu g$ represents the effective atmospheric height. We adopted the typical atmospheric mean molecular mass $\mu$ to be 2.3 amu for sub-Neptune planets \citep{Demory2020}. Assuming the atmospheric temperature $T$ to be the equilibrium temperature $T_{\rm eq}$, we obtained an amplitude of $243^{+164}_{-75}$ ppm\footnote{The large error bar mainly comes from the uncertainty of the planet mass $M_{p}.$} of \tar b, which is above the noise floor level 50 ppm of \jwst\ for MIRI LRS ($ \lambda= 5.0 - 11\mu {\rm m}$) observations \citep{Greene2016}. 

We further computed the Transmission Spectroscopy Metric (TSM; \citealt{Kempton2018}) of \tar b to be $126^{+168}_{-54}$. \cite{Kempton2018} recommended that planets with TSM$>90$ and $1.5<R_{p}<10\ R_{\oplus}$ are high-quality atmospheric characterization targets. Combined with the two aspects above, we regarded \tar b as an attractive source for further atmosphere composition analysis. 

\subsection{Prospects on Future Follow-up Observations}
Since the current RV datasets only enable a $\sim 2\sigma$ mass constraint, here we suggest that future RVs measurements of \tar\ are needed to break the degeneracy between the planet mass $M_{p}$ and the mean molecular weight $\mu$ \citep{Seager2000,Seager2009} in any atmospheric characterization studies. Additional RVs would also be crucial to search for outer long period non-transiting cold giants (See the next subsection). 

Given the brightness of \tar\ ($V=8.7$~mag), most high resolution optical spectroscopy facilities like the Planet Finder Spectrograph (PFS; \citealt{Crane2006,Crane2008,Crane2010}) can achieve high SNR and reach the 1~m/s precision. In addition, the expected rotation timescale of \tar\ is well separated from the planet orbital period, making it possible to smooth out the stellar activity effects in RVs with high cadence observations \citep{Morales2016,Gan2020b}.

\subsection{Additional planet in the \tar\ system?}
\cite{Zhu2018} suggested a higher probability of detecting cold Jupiters around the hosts of small planets (planets with mass/radius between Earth and Neptune) compared with other field stars, which is also supported by the recent observational results from \cite{Bryan2019}. As we have a long time baseline of HARPS observations ($\sim 3000$~d), we first looked for possible periodic signals on the residuals after subtracting the best-fit model for \tar b using GLS. However, we did not identify any significant peaks between 100 d and 1500 d with FAP<0.1\%. We then ran another BM+1pl fit to blindly search the potential cold giant planets with period between the aforementioned range. We fit the Keplerian signals of \tar b and another potential outer cold planet simultaneously. By adopting the same prior settings for other parameters as the BM+1pl model in Section \ref{rv}, we obtained $\Delta \ln Z=\ln Z_{\rm BM+1pl}-\ln Z_{\rm BM}=3$, which ruled out the existence of an outer gas giant planet with mass down to Saturn mass and period up to roughly 4 years based on the current data.

\subsection{Comparison with Palatnick et al. 2021}\label{discrepancy}
We note that during the writing of this manuscript, \cite{Palatnick2021} validated this system using \tess\ and HARPS archival-only data (56 in total). The RV semi-amplitude of \tar\ reported by \cite{Palatnick2021} is $4.9^{+0.9}_{-1.0}$ m/s, which is about $1.7\sigma$ larger than our measurement $2.7^{+1.3}_{-1.3}$ m/s. The discrepancy is caused by the contribution from the three recent, additional HARPS RV points taken between 2019 Aug. 16th and 2019 Aug. 20th in this work. All of the three newly acquired HARPS spectra have high SNR (105.9, 110.3 and 95.6, respectively). Without including these HARPS RV data, as in \cite{Palatnick2021}, we derived a similar RV semi-amplitude of $K_{b}=5.2^{+1.1}_{-1.2}$ m/s by fitting a Keplerian model plus a linear RV slope $\dot{\gamma}$ using the same prior settings described in Section \ref{rv}\footnote{For the RV slope term, we adopted a uniform prior with an initial guess of 0 $\rm m\ s^{-1}d^{-1}$.}. The best-fit $\dot{\gamma}$ is $\rm -3.3^{+0.9}_{-0.9}\ m\ s^{-1}yr^{-1}$, consistent with the posterior value of $\rm -3.2^{+0.8}_{-0.7}\ m\ s^{-1}yr^{-1}$ reported in \cite{Palatnick2021}. However, after including the three recent HARPS RV points\footnote{We removed one archival HARPS measurement whose BIS and FWHM were outliers (see Section \ref{harps} for more detail) so the final RV data number used in this model is 58.}, we found the model prefers a null RV slope ($\dot{\gamma}=-0.1^{+0.6}_{-0.8}\ {\rm m\ s^{-1}yr^{-1}}$) and the RV semi-amplitude decreases to $K_{b}=2.3^{+1.2}_{-1.2}$ m/s. We show both models in Figures \ref{comparison_v1} and \ref{comparison_v2}. This difference is reasonable since the stellar activity plays a role in the Doppler signals, as stated in Section \ref{rotation}, which biases the result of \cite{Palatnick2021}. Thus a Keplerian+RV slope model may not be able to explain the total HARPS data set. We emphasize here that our results are still consistent within $1.7\sigma$, and more RV data are needed to deal with the stellar activity and measure the mass of the planet more accurately.

\section{Summary and Conclusions}\label{conclusion}
In this paper, we characterize the \tar\ planetary system using both space and ground-based photometric data from \tess\ and LCO as well as the spectroscopic data from HARPS and \textsc{Minerva}-Australis. Our models reveal that \tar b is a warm sub-Neptune hosted by a nearby solar twin with an orbital period of $17.47$~d, a radius of $3.53_{-0.11}^{+0.13}\ R_{\oplus}$ and a mass of $11.2_{-5.4}^{+5.4}\ M_{\oplus}$, with a $3\sigma$ upper limit of $27.4\ M_{\oplus}$ (see Figure \ref{mr}). Taken together, the resulting planetary bulk density of $\rm 1.4^{+0.9}_{-0.8}\ g\ cm^{-3}$, implies that an extended atmosphere is likely present, making this system an excellent candidate for transmission spectroscopic follow-up. The line-by-line differential spectroscopic analysis shows that \tar\ does not show a similar depletion in the abundance of refractory elements as our Sun. The lack of a Solar refractory depletion could plausibly be linked to \tar's lack of known giant planets (following the gap-opening hypothesis) or a history of planetary migration and stellar infall, especially if the planet reported here did not form in-situ (following the planet accretion hypothesis).

\section*{Affiliations}
$^{1}$Department of Astronomy and Tsinghua Centre for Astrophysics, Tsinghua University, Beijing 100084, China\\
$^{2}$\flatiron\\
$^{3}$\USP\\
$^{4}$National Astronomical Observatories, Chinese Academy of Sciences, 20A Datun Road, Chaoyang District, Beijing 100012, China\\
$^{5}$Department of Physics and Astronomy, Vanderbilt University, 6301 Stevenson Center Ln., Nashville, TN 37235, USA\\
$^{6}$Department of Physics, Fisk University, 1000 17th Avenue North, Nashville, TN 37208, USA\\
$^{7}$NASA Ames Research Center, Moffett Field, CA 94035, USA\\
$^{8}$Department of Physics, Engineering and Astronomy, Stephen F. Austin State University, 1936 North St, Nacogdoches, TX 75962, USA\\
$^{9}$University of Southern Queensland, Centre for Astrophysics, West Street, Toowoomba, QLD 4350 Australia\\
$^{10}$Astrophysics Group, Keele University, Staffordshire, ST5 5BG, UK\\
$^{11}$\harvard\\
$^{12}$\MIT\\
$^{13}$Department of Earth, Atmospheric and Planetary Sciences, Massachusetts Institute of Technology, Cambridge, MA 02139, USA\\
$^{14}$Department of Aeronautics and Astronautics, MIT, 77 Massachusetts Avenue, Cambridge, MA 02139, USA\\
$^{15}$Department of Astrophysical Sciences, Princeton University, 4 Ivy Lane, Princeton, NJ 08544, USA\\
$^{16}$Department of Astronomy, University of Florida, 211 Bryant Space Science Center, Gainesville, FL, 32611, USA\\
$^{17}$NASA Goddard Space Flight Center, 8800 Greenbelt Road, Greenbelt, MD 20771, USA\\
$^{18}$University of Maryland, Baltimore County, 1000 Hilltop Circle, Baltimore, MD 21250, USA\\
$^{19}$\chicago\\
$^{20}$Department of Astronomy, The University of Texas at Austin, TX 78712, USA\\
$^{21}$Cerro Tololo Inter-American Observatory, Casilla 603, La Serena, Chile\\
$^{22}$Department of Physics \& Astronomy, University of Kansas, 1082 Malott,1251 Wescoe Hall Dr., Lawrence, KS 66045, USA\\
$^{23}$Max Planck Institute for Astronomy, Konigstuhl 17, 69117 Heidelberg, Germany\\
$^{24}$Department of Physics \& Astronomy, Swarthmore College, Swarthmore PA 19081, USA\\
$^{25}$Department of Earth and Planetary Sciences, University of California, Riverside, CA 92521, USA\\
$^{26}$Department of Physics and Astronomy, University of Louisville, Louisville, KY 40292, USA\\
$^{27}$Department of Physics and Astronomy, The University of North Carolina at Chapel Hill, Chapel Hill, NC 27599-3255, USA\\
$^{28}$NCCR/PlanetS, Centre for Space \& Habitability, University of
Bern, Bern, Switzerland\\
$^{29}$George Mason University, 4400 University Drive MS 3F3, Fairfax, VA 22030, USA\\
$^{30}$Patashnick Voorheesville Observatory, Voorheesville, NY 12186, USA\\
$^{31}$Space Telescope Science Institute, 3700 San Martin Drive, Baltimore, MD 21218, USA\\
$^{32}$SETI Institute, 189 Bernardo Ave, Suite 200, Mountain View, CA 94043, USA\\
$^{33}$INAF - Osservatorio Astronomico di Padova, vicolo dell'Osservatorio 5, 35122, Padova, Italy\\
$^{34}$Exoplanetary Science at UNSW, School of Physics, UNSW Sydney, NSW 2052, Australia\\
$^{35}$Tsinghua International School, Beijing 100084, China\\
$^{36}$Stanford Online High School, 415 Broadway Academy Hall, Floor 2, 8853, Redwood City, CA 94063, USA\\
$^{37}$School of Astronomy and Space Science, Key Laboratory of Modern Astronomy and Astrophysics in Ministry of Education, Nanjing University, Nanjing 210046, Jiangsu, China\\
\section*{Acknowledgements}

We thank Jennifer Burt and Chelsea X. Huang for bringing this TOI to our attention, and Oscar Barrag\'{a}n, Trevor David, Annelies Mortier and Andrew Vanderburg for useful discussions.
This work is partly supported by the National Science Foundation of China (Grant No. 11390372 and 11761131004 to SM and TG). 
This research uses data obtained through the Telescope Access Program (TAP), which has been funded by the TAP member institutes.
JM thanks FAPESP (2018/04055-8).
HZ acknowledges NSFC: 12073010, which supports the collaboration with the \textsc{Minerva}-Australis team. 
This work has been carried out within the framework of the National Centre of Competence in Research PlanetS supported by the Swiss National Science Foundation. Hugh Osborn acknowledges the financial support of the SNSF.
Funding for the TESS mission is provided by NASA's Science Mission directorate. 
This work is based on observations collected at the European Southern Observatory under ESO programmes 188.C-0265 and 0100.D-0444. 
This work has made use of data from the European Space Agency (ESA) mission
{\it Gaia} (\url{https://www.cosmos.esa.int/gaia}), processed by the {\it Gaia} Data Processing and Analysis Consortium (DPAC,
\url{https://www.cosmos.esa.int/web/gaia/dpac/consortium}). Funding for the DPAC has been provided by national institutions, in particular the institutions participating in the {\it Gaia} Multilateral Agreement.
We acknowledge the use of \tess\ Alert data from pipelines at the \tess\ Science Office and at the \tess\ Science Processing Operations Center. 
Resources supporting this work were provided by the NASA High-End Computing (HEC) Program through the NASA Advanced Supercomputing (NAS) Division at Ames Research Center for the production of the SPOC data products.
\textsc{Minerva}-Australis is supported by Australian Research Council LIEF Grant LE160100001, Discovery Grant DP180100972, Mount Cuba Astronomical Foundation, and institutional partners University of Southern Queensland, UNSW Sydney, MIT, Nanjing University, George Mason University, University of Louisville, University of California Riverside, University of Florida, and The University of Texas at Austin. We respectfully acknowledge the traditional custodians of all lands throughout Australia, and recognise their continued cultural and spiritual connection to the land, waterways, cosmos, and community. We pay our deepest respects to all Elders, ancestors and descendants of the Giabal, Jarowair, and Kambuwal nations, upon whose lands the \textsc{Minerva}-Australis facility at Mt Kent is situated.
Some of the observations in the paper made use of the High-Resolution Imaging instrument Zorro at Gemini-South). Zorro was funded by the NASA Exoplanet Exploration Program and built at the NASA Ames Research Center by Steve B. Howell, Nic Scott, Elliott P. Horch, and Emmett Quigley.
This research has made use of the Exoplanet Follow-up Observation Program website, which is operated by the California Institute of Technology, under contract with the National Aeronautics and Space Administration under the Exoplanet Exploration Program. 
This paper includes data collected by the \tess\ mission, which are publicly available from the Mikulski Archive for Space Telescopes\ (MAST). 
This work made use of \texttt{tpfplotter} by J. Lillo-Box (publicly available in www.github.com/jlillo/tpfplotter), which also made use of the python packages \texttt{astropy}, \texttt{lightkurve}, \texttt{matplotlib} and \texttt{numpy}.
This research made use of \textsf{exoplanet} \citep{exoplanet:exoplanet} and its dependencies \citep{exoplanet:astropy13, exoplanet:astropy18, exoplanet:pymc3, exoplanet:theano}.
This research made use of observations from the LCO network, WASP-South and ESO: 3.6m\ (HARPS). 

\section*{Data Availability}

This paper includes photometric data collected by the \tess\ mission and LCOGT, which is publicly available in ExoFOP, at \url{https://exofop.ipac.caltech.edu/tess/target.php?id=320004517}. All spectroscopy data underlying this article are listed in the appendix.



\bibliographystyle{mnras}
\bibliography{planet} 




\appendix

\section{All RVs and stellar activity indicators of HARPS and \textsc{Minerva}-Australis}

\begin{table*}
    \centering
    \caption{HARPS RVs and stellar activity indicators}
    \begin{tabular}{ccccccccccc}
        \hline\hline
        Time (BJD)   &RV (m/s)  &RVerr (m/s)     &CRX &CRXerr &DLW  &DLWerr &BIS &FWHM &$S_{\rm HK}$ &$S_{\rm HK}$err   \\\hline
2455847.536 &14.37 &1.53 &-1.16 &12.18 &1.99 &2.49 &-0.025 &7.091 &0.1821 &0.0014\\
2455850.515 &2.02 &0.82 &-10.75 &6.45 &-1.97 &1.27 &-0.016 &7.079 &0.191 &0.0008\\
2455851.518 &-0.3 &0.71 &1.63 &5.75 &-7.83 &1.18 &-0.015 &7.068 &0.1901 &0.0007\\
2455852.505 &-1.39 &1.22 &3.17 &9.84 &-10.34 &1.64 &-0.019 &7.062 &0.1897 &0.0009\\
2456042.793 &4.35 &1.15 &11.92 &9.15 &-10.75 &1.47 &-0.024 &7.073 &0.1852 &0.001\\
2456043.878 &3.59 &1.21 &-3.85 &9.74 &-8.64 &1.55 &-0.019 &7.074 &0.1854 &0.001\\
2456045.888 &15.57 &1.37 &-25.38 &10.55 &-2.59 &2.25 &-0.018 &7.078 &0.1894 &0.0012\\
2456046.939 &10.58 &0.86 &-17.16 &6.6 &7.91 &1.27 &-0.017 &7.082 &0.1909 &0.0009\\
2456048.942 &7.96 &0.97 &-8.24 &7.75 &19.62 &1.82 &-0.015 &7.075 &0.1944 &0.0009\\
2456162.592 &10.61 &2.11 &-14.78 &16.61 &6.19 &3.48 &-0.014 &7.087 &0.1872 &0.0018\\
2456164.651 &7.01 &1.49 &-10.12 &11.79 &-2.52 &2.14 &-0.017 &7.082 &0.1972 &0.0015\\
2456165.633 &9.13 &1.01 &-8.8 &8.1 &-1.67 &1.44 &-0.019 &7.072 &0.1965 &0.0009\\
2456378.907 &-7.67 &0.82 &14.47 &6.39 &-0.23 &1.1 &-0.019 &7.073 &0.1892 &0.0008\\
2456484.744 &-3.47 &1.81 &35.43 &13.85 &-0.01 &2.17 &-0.026 &7.09 &0.1769 &0.0014\\
2456485.724 &-2.08 &1.03 &19.75 &7.82 &-3.14 &1.34 &-0.021 &7.081 &0.1912 &0.0008\\
2456486.702 &1.27 &0.93 &23.83 &6.83 &0.36 &0.89 &-0.024 &7.084 &0.1945 &0.0007\\
2456487.706 &2.01 &0.96 &27.73 &6.86 &-0.97 &0.93 &-0.022 &7.081 &0.195 &0.0007\\
2456488.727 &3.66 &0.98 &29.05 &6.91 &2.78 &1.23 &-0.022 &7.087 &0.1939 &0.0009\\
2456489.694 &7.93 &0.99 &-12.21 &7.79 &5.59 &1.36 &-0.011 &7.09 &0.1934 &0.0009\\
2456490.7 &1.49 &0.92 &-16.04 &7.1 &2.04 &1.19 &-0.018 &7.081 &0.1928 &0.0008\\
2456557.59 &6.92 &1.65 &12.62 &12.95 &4.51 &3.16 &-0.017 &7.087 &0.1926 &0.0018\\
2456558.572 &3.15 &0.98 &2.53 &7.89 &-0.06 &1.18 &-0.02 &7.073 &0.191 &0.0008\\
2456559.583 &7.07 &1.06 &-11.08 &8.36 &-2.82 &1.43 &-0.02 &7.077 &0.1885 &0.001\\
2456560.578 &3.09 &0.83 &-13.03 &6.48 &0.53 &1.15 &-0.023 &7.071 &0.189 &0.0008\\
2456850.714 &-0.38 &1.23 &-6.28 &9.74 &6.38 &1.74 &-0.017 &7.072 &0.1932 &0.0012\\
2456851.725 &2.3 &1.04 &-7.54 &8.29 &5.97 &1.36 &-0.016 &7.076 &0.1869 &0.001\\
2456852.73 &1.74 &1.07 &-19.51 &8.19 &-1.3 &1.71 &-0.021 &7.075 &0.1939 &0.001\\
2456853.794 &9.94 &1.93 &-6.63 &15.39 &5.63 &2.25 &-0.018 &7.088 &0.173 &0.0014\\
2456855.725 &8.36 &1.11 &-17.74 &8.68 &1.32 &1.45 &-0.02 &7.083 &0.1916 &0.0009\\
2456856.715 &4.03 &0.83 &-13.81 &6.47 &3.66 &1.27 &-0.015 &7.084 &0.1947 &0.0009\\
2456904.574 &5.07 &0.92 &-21.63 &6.78 &-1.63 &1.28 &-0.023 &7.074 &0.1872 &0.0009\\
2456906.573 &5.58 &1.58 &6.81 &12.49 &1.74 &2.79 &-0.018 &7.079 &0.1792 &0.0015\\
2456907.587 &4.38 &1.24 &-13.04 &9.71 &1.04 &1.76 &-0.022 &7.076 &0.194 &0.0012\\
2456961.505 &11.57 &2.43 &-7.7 &19.4 &5.59 &3.13 &-0.028 &7.083 &0.1818 &0.0019\\
2456965.501 &-3.57 &0.75 &1.78 &6.02 &-0.66 &0.97 &-0.021 &7.082 &0.1885 &0.0007\\
2456966.521 &-1.39 &1.01 &0.58 &8.2 &-2.71 &1.35 &-0.023 &7.079 &0.1893 &0.0009\\
2457226.716 &10.4 &0.96 &-3.69 &7.98 &4.29 &1.1 &-0.004 &7.1 &0.1934 &0.0007\\
2457227.703 &10.21 &1.11 &-4.51 &9.22 &4.28 &0.9 &-0.007 &7.104 &0.1953 &0.0007\\
2457228.698 &4.03 &1.22 &-6.97 &10.15 &4.88 &1.3 &-0.004 &7.104 &0.196 &0.001\\
2457229.711 &4.87 &1.07 &0.36 &9.06 &4.28 &0.79 &-0.004 &7.1 &0.1927 &0.0006\\
2457230.698 &1.2 &0.92 &1.46 &7.79 &2.64 &0.83 &-0.004 &7.095 &0.1891 &0.0006\\
2457232.67 &1.54 &1.52 &-7.21 &12.74 &3.76 &1.83 &-0.002 &7.104 &0.1861 &0.0012\\
2457283.549 &-6.69 &1.56 &0.72 &12.94 &-5.74 &2.0 &-0.004 &7.088 &0.1875 &0.0011\\
2457284.601 &-5.29 &1.05 &7.82 &8.72 &-2.45 &1.1 &-0.009 &7.09 &0.1889 &0.0008\\
2457507.88 &2.41 &1.5 &-1.29 &12.65 &7.57 &1.16 &-0.004 &7.117 &0.1824 &0.0006\\
2457587.718 &1.44 &1.06 &4.99 &8.79 &-4.88 &1.24 &-0.007 &7.087 &0.1807 &0.0008\\
2457588.767 &1.86 &1.02 &-9.51 &8.38 &-5.32 &0.91 &0.036 &7.139 &0.1791 &0.0008\\
2457588.78 &2.33 &1.08 &0.07 &8.89 &-4.64 &1.46 &-0.012 &7.082 &0.1797 &0.0009\\
2457664.611 &3.37 &0.96 &-9.33 &7.8 &-5.5 &0.97 &-0.005 &7.095 &0.1857 &0.0008\\
2457665.549 &-2.78 &1.04 &9.86 &8.57 &-3.84 &1.14 &-0.006 &7.096 &0.184 &0.0008\\
2457682.499 &-1.45 &0.85 &9.3 &6.97 &-3.19 &1.08 &-0.013 &7.09 &0.186 &0.0007\\
2457683.568 &-3.81 &0.97 &3.03 &8.04 &-7.46 &1.26 &-0.018 &7.088 &0.1875 &0.001\\
2458047.504 &0.16 &0.98 &-2.6 &8.15 &-3.8 &0.67 &-0.009 &7.088 &0.184 &0.0006\\
2458711.619 &15.76 &1.18 &4.2 &9.88 &2.83 &1.19 &-0.008 &7.1 &0.204 &0.0008\\
2458713.76 &13.1 &1.14 &9.7 &9.52 &5.74 &1.38 &-0.006 &7.102 &0.1998 &0.001\\
2458715.714 &8.67 &1.08 &-10.43 &9.04 &2.9 &1.14 &-0.009 &7.103 &0.2021 &0.0009\\
        \hline\hline 
    \end{tabular}
    \label{harpsai}
\end{table*}

\begin{table}
    \centering
    \caption{\textsc{Minerva}-Australis RVs of 4 telescopes: MA, MB, MC and MD}
    \begin{tabular}{cccc}
        \hline\hline
        Time(BJD)   &RV(m/s)  &RVerr(m/s)     &Instrument  \\\hline
        2458961.205 &13.51 &8.03 &MA\\
2458964.276 &-5.64 &8.17 &MA\\
2458966.254 &0.3 &8.35 &MA\\
2458980.084 &-6.58 &8.67 &MA\\
2458981.194 &20.78 &7.4 &MA\\
2458959.285 &-17.62 &5.0 &MB\\
2458961.205 &11.8 &6.36 &MB\\
2458964.276 &-4.74 &5.22 &MB\\
2458966.254 &-1.2 &5.29 &MB\\
2458971.214 &1.2 &8.62 &MB\\
2458971.236 &-17.36 &8.93 &MB\\
2458973.184 &19.1 &5.2 &MB\\
2458973.206 &13.31 &5.17 &MB\\
2458959.285 &-34.61 &6.39 &MC\\
2458961.205 &-14.81 &7.23 &MC\\
2458964.276 &-28.45 &6.54 &MC\\
2458966.254 &-41.26 &6.91 &MC\\
2458973.184 &8.17 &6.09 &MC\\
2458973.206 &9.67 &6.35 &MC\\
2458974.308 &1.66 &7.46 &MC\\
2458980.084 &-0.27 &8.1 &MC\\
2458980.106 &20.03 &8.04 &MC\\
2458981.172 &7.12 &6.76 &MC\\
2458981.194 &14.62 &6.66 &MC\\
2458995.32 &-31.32 &8.98 &MC\\
2458998.047 &-11.82 &6.91 &MC\\
2458998.069 &9.99 &6.86 &MC\\
2459002.078 &-13.7 &7.22 &MC\\
2458995.32 &-20.95 &5.55 &MD\\
2458998.047 &6.44 &6.23 &MD\\
2458998.069 &0.0 &5.57 &MD\\
2459002.078 &-3.96 &5.75 &MD\\
2459002.099 &3.82 &7.18 &MD\\
        \hline\hline 
    \end{tabular}
\end{table}

\section{BM for HARPS and \textsc{Minerva}}
\begin{figure}
\centering
\includegraphics[width=0.5\textwidth]{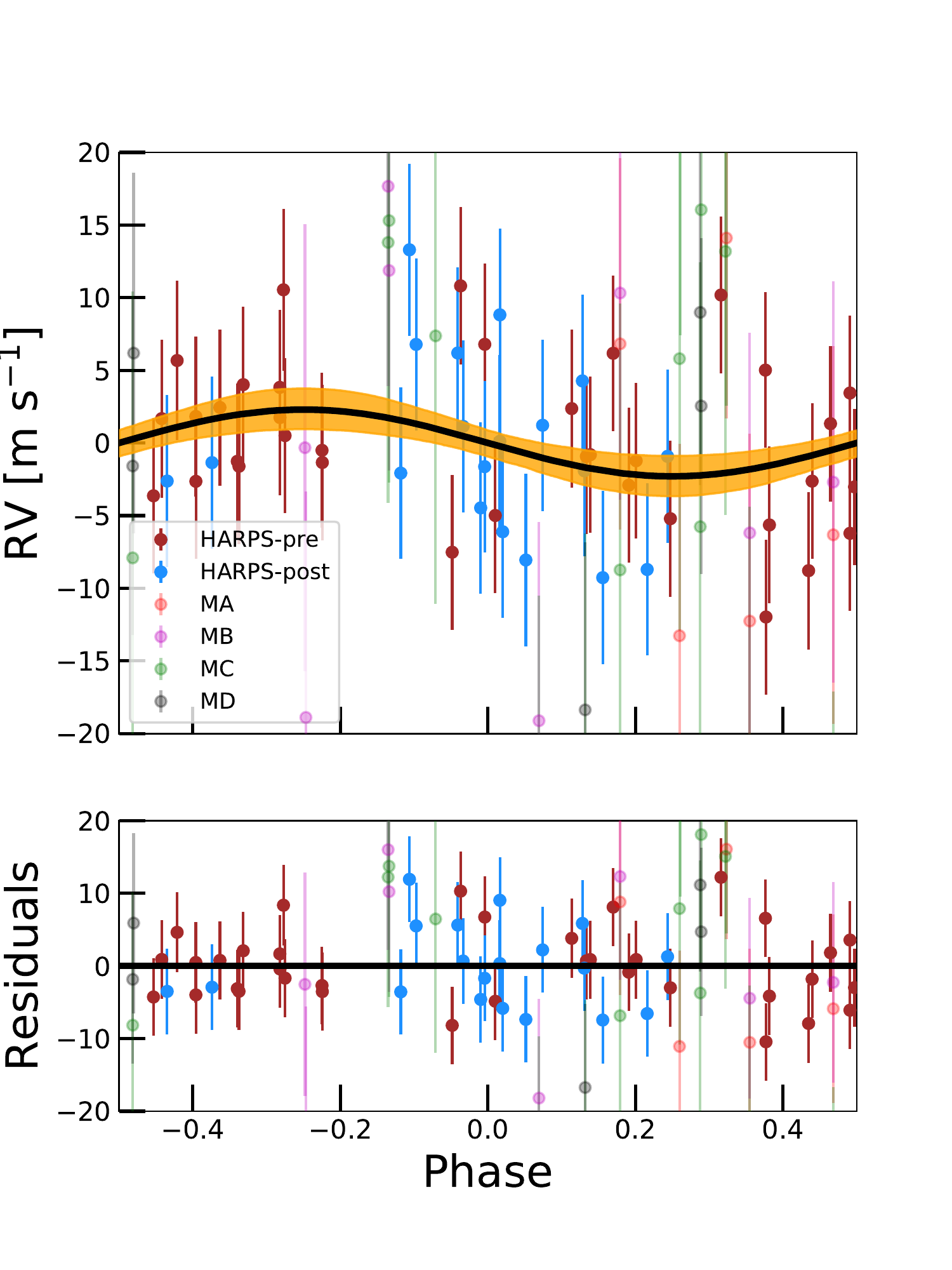}
\caption{The phase-folded HARPS and \textsc{Minerva} RVs of \tar. The best-fit base model is shown as a black solid line. The orange shaded region represents the $1\sigma$ confidence interval of the model. Residuals are plotted below.} 
\label{allrv}
\end{figure}

\section{Model Comparison}
\begin{figure*}
\centering
\includegraphics[width=0.99\textwidth]{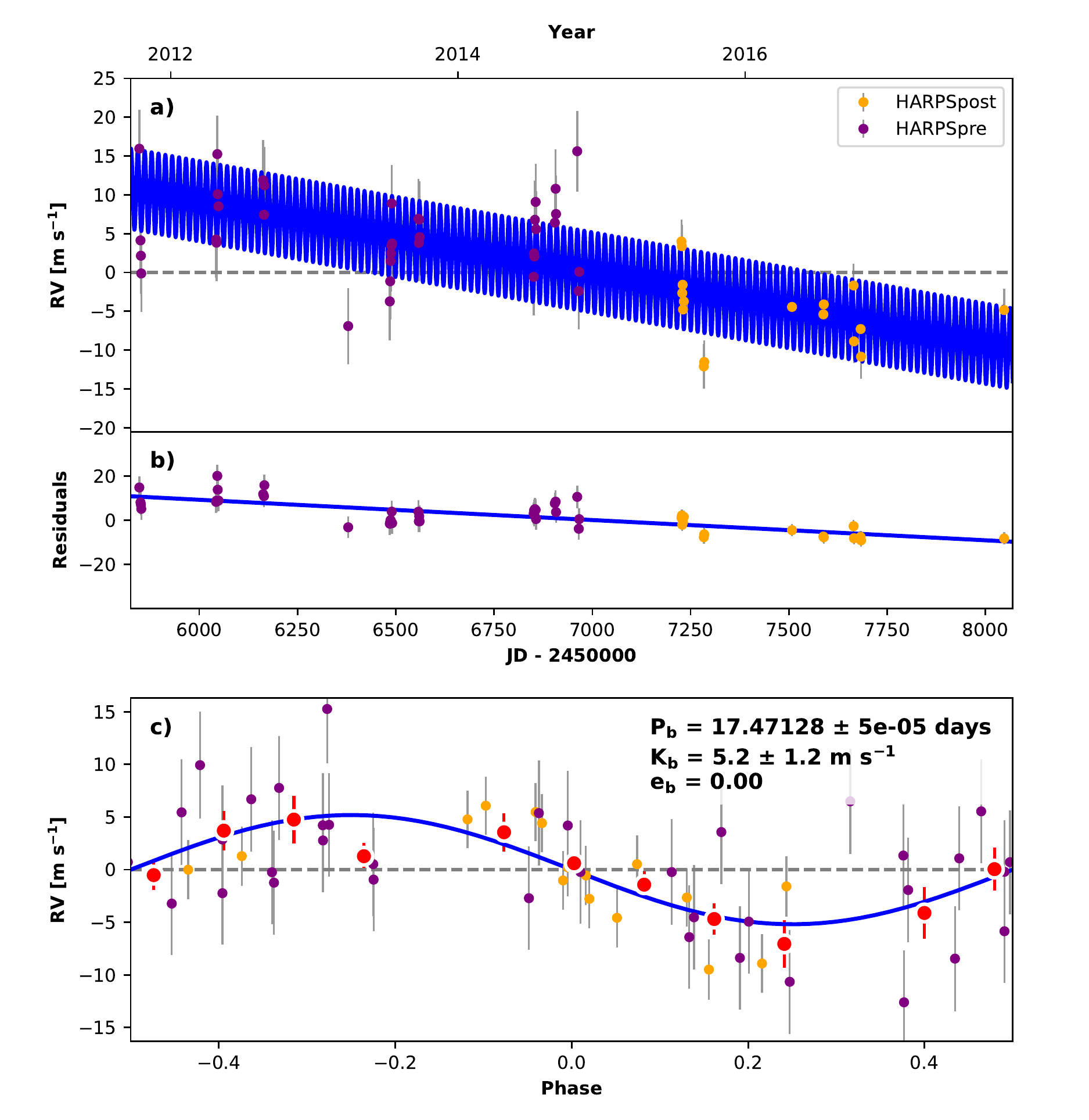}
\caption{HARPS archive-only data used in \citet{Palatnick2021} and the best-fit model. The top panel shows the full RV time series and residuals. The phase-folded RV data are presented in the bottom panel. The red points are the binned RVs.} 
\label{comparison_v1}
\end{figure*}

\begin{figure*}
\centering
\includegraphics[width=0.99\textwidth]{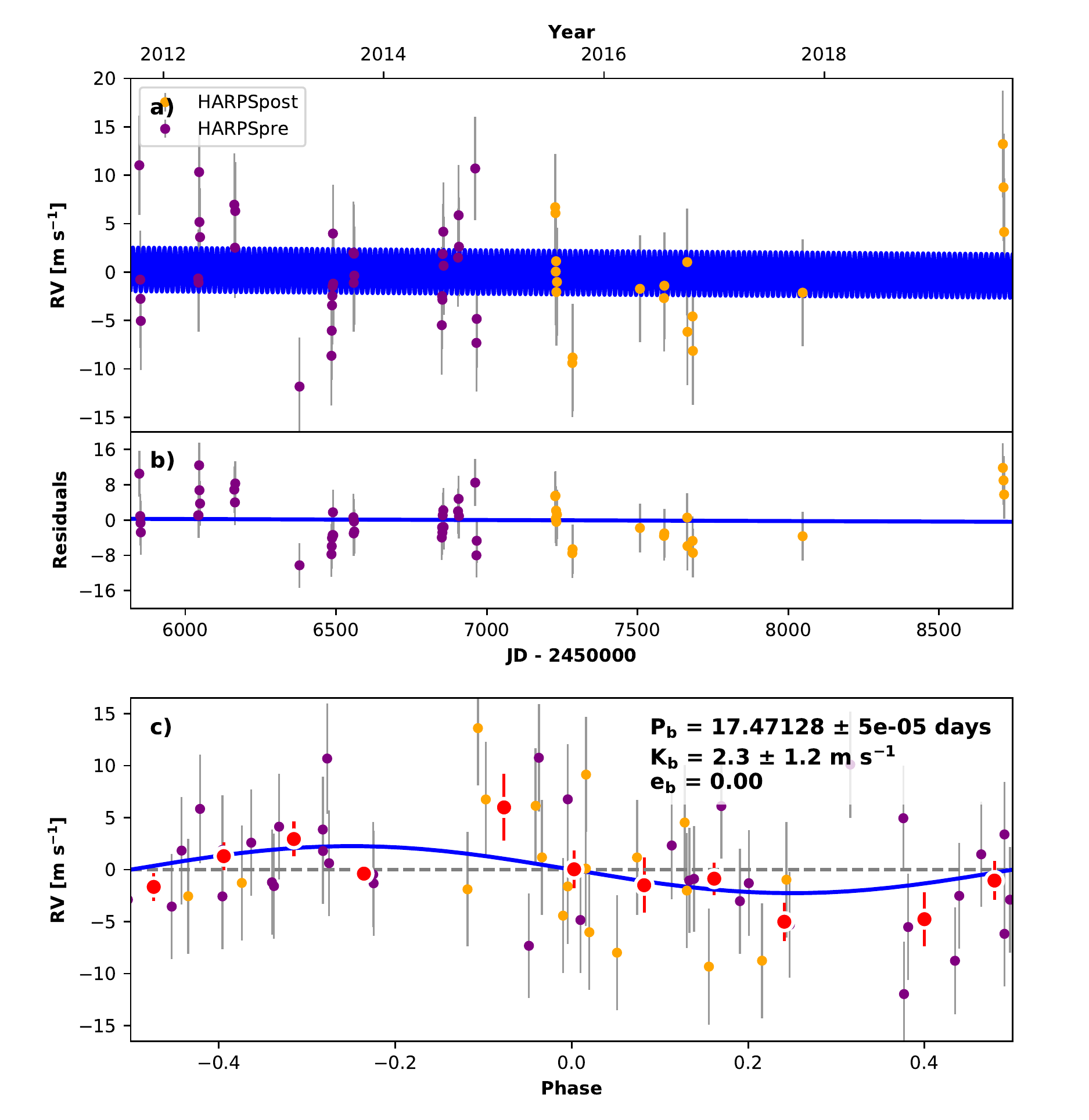}
\caption{HARPS archive-only data plus three additional RV points along with the best-fit model. The top panel shows the full RV time series and residuals. The phase-folded RV data are presented in the bottom panel. The red points are the binned RVs. After including the new HARPS data, the model prefers a null RV slope (see Section \ref{discrepancy}).} 
\label{comparison_v2}
\end{figure*}


\bsp	
\label{lastpage}
\end{document}